\documentclass[prd,preprintnumbers,floatfix,
nofootinbib,superscriptaddress]{revtex4}
\usepackage{float}
\usepackage{nicefrac}
\usepackage{slashed}
\usepackage{mathtools}
\usepackage{amsfonts} % AMS
\usepackage{amssymb} % AMS
\usepackage{amsmath} % AMS
\usepackage{graphicx} % Include figure files
\usepackage{subfigure} % Include figure files
\usepackage{array} % array
\usepackage{dcolumn} % Align table columns on decimal point
\usepackage{bm} % bold math
\usepackage{latexsym} % latex symbols
\usepackage{longtable} % long tables
\usepackage{hyperref} % hypertext links 
\usepackage{verbatim}
\usepackage{epsfig}
\usepackage{xcolor}
\usepackage{color}
\usepackage{cancel}
\usepackage{braket}
\usepackage{xspace}
\usepackage{soul}
\usepackage{physics} % shorthand latex for pdvs
\usepackage{outlines}
\newcommand{\Ac}[0]{{\mathcal{A}}}
\newcommand{\Bc}[0]{{\mathcal{B}}}
\newcommand{\Cc}[0]{{\mathcal{C}}}

\newcommand{\Gc}[0]{{\mathcal{G}}}
\newcommand{\Hc}[0]{{\mathcal{H}}}

\newcommand{\Jc}[0]{{\mathcal{J}}}
\newcommand{\Kc}[0]{{\mathcal{K}}}

\newcommand{\Mc}[0]{{\mathcal{M}}}

\newcommand{\Tc}[0]{{\mathcal{T}}}
\newcommand{\Wc}[0]{{\mathcal{W}}}
\newcommand{\Yc}[0]{{\mathcal{Y}}}

\newcommand{\Ab}[0]{{\mathbf{A}}}
\newcommand{\Bb}[0]{{\mathbf{B}}}
\newcommand{\Hb}[0]{{\mathbf{H}}}
\newcommand{\Ib}[0]{{\mathbf{I}}}
\newcommand{\Tb}[0]{{\mathbf{T}}}
\newcommand{\Wb}[0]{{\mathbf{W}}}

\newcommand{\Ob}[0]{{\mathbf{O}}}

\newcommand{\df}[0]{{\rm{df}}}
\newcommand{\diff}[0]{{\rm{d}}}
\newcommand{\nn}[0]{\nonumber}

\renewcommand{\vec}[1]{\mathbf{#1}}

\begin{document}
\title{
  Two-current transition amplitudes with two-body final states
}

%%%%%%%%%%%%%%%%%%%%%
\author{Keegan H. Sherman}
\email[e-mail: ]{ksher004@odu.edu}
\affiliation{Department of Physics, Old Dominion University, Norfolk, Virginia 23529, USA}
%%%%%%%%%%%%%%%%%%%%%
\author{Felipe G. Ortega-Gama}
\email[e-mail: ]{fgortegagama@email.wm.edu}
\affiliation{Thomas Jefferson National Accelerator Facility, 12000 Jefferson Avenue, Newport News, Virginia 23606, USA}
\affiliation{Department of Physics, William \& Mary, Williamsburg, Virginia 23187, USA}
%%%%%%%%%%%%%%%%%%%%%
\author{Ra\'ul A.~Brice\~no}
\email[e-mail: ]{rbriceno@jlab.org}
\affiliation{Thomas Jefferson National Accelerator Facility, 12000 Jefferson Avenue, Newport News, Virginia 23606, USA}
\affiliation{ Department of Physics, Old Dominion University, Norfolk, Virginia 23529, USA}
%%%%%%%%%%%%%%%%%%%%%
\author{Andrew W. Jackura}
\email[e-mail: ]{ajackura@odu.edu}
\affiliation{Thomas Jefferson National Accelerator Facility, 12000 Jefferson Avenue, Newport News, Virginia 23606, USA}
\affiliation{ Department of Physics, Old Dominion University, Norfolk, Virginia 23529, USA}
%%%%%%%%%%%%%%%%%%%%%
\preprint{JLAB-THY-22-3552}
\date{\today}
\begin{abstract}
We derive the on-shell form of amplitudes containing two external currents with a single hadron in the initial state and two hadrons in the final state, denoted as $1+\mathcal{J}\to 2+\mathcal{J}$. This class of amplitude is relevant in precision tests of the Standard Model as well as for exploring the structure of excited states in the QCD spectrum. We present a model-independent  description of the amplitudes where we sum to all orders in the strong interaction. From this analytic form we are able to extract transition and elastic resonance form factors consistent with previous work as well as a novel Compton-like amplitude coupling a single particle state to a resonance. The results also hold for reactions where the one-particle state is replaced with the vacuum, namely $\mathcal{J}\to 2+\mathcal{J}$ amplitudes. We also investigate constraints placed upon the formalism for the case of a conserved vector current in the form of the Ward-Takahashi identity. The formalism presented here is valid for currents of arbitrary Lorentz structure and quantum numbers with spinless hadrons where any number of two-particle intermediate channels may be open. When combined with the appropriate finite-volume framework, this work facilitates the extraction of physical observables from this class of amplitudes via lattice QCD calculations.
\end{abstract}

% \keywords{}
%
\maketitle

%%%%%%%%%%%%%%%%%%%%%%%%%%%%%%%%%%%%%%%%%%%
%  Section: Introduction
%%%%%%%%%%%%%%%%%%%%%%%%%%%%%%%%%%%%%%%%%%%
\section{Introduction \label{sec:intro}}
  
Quantifying non-perturbative Quantum Chromodynamic (QCD) contributions to electroweak interactions of hadronic processes remains an on-going challenge in modern nuclear and particle physics. Several outstanding problems lie in descriptions of hadronic transitions involving timelike separated external currents, including computing hadronic structure functions and precision tests of the Standard Model. An important example in the context of precision measurements is the anomalous magnetic moment of the muon, $a_{\mu}$, where presently tension persists between the theoretical prediction and the experimental measurement~\cite{Aoyama:2020ynm}. 
Attempts have been made using phenomenological analyses~\cite{Colangelo:2017fiz,Hoferichter:2019nlq,Danilkin:2018qfn,Danilkin:2019opj, Gasser:2005ud,Gasser:2006qa,Colangelo:2018mtw,Keshavarzi:2019abf,Davier:2019can,Ananthanarayan:2018nyx,Benayoun:2019zwh} and lattice QCD~\cite{Blum:2014oka,Blum:2015gfa,Blum:2016lnc,Green:2015sra,Asmussen:2018oip} to determine the contributions that dominate the theoretical uncertainty of $a_{\mu}$, which are the hadronic vacuum polarization (HVP) and hadronic light-by-light (HLbL) tensors.
At leading order in Quantum Electrodynamics (QED), the HVP and HLbL tensors can be written in terms of hadronic matrix elements of the QED current $\mathcal{J}^\mu$ of the form $\langle 0| T\left[\prod_{j=1}^2\mathcal{J}^{\mu_j}(x_j)   \right]  |0\rangle$ and $\langle 0|T\left[\prod_{j=1}^4\mathcal{J}^{\mu_j}(x_j)   \right]|0\rangle$, respectively~\footnote{The $\mu_j$ superscripts denote the Lorentz index of the current, and $x_j$ are the corresponding spacetime points.}. 
A promising effort to determine the light-by-light amplitude, which is the hardest to constrain, is to use a dispersive representation of this amplitude, in terms of, among other things, $\gamma^\star \gamma^\star \to \pi\pi, K\overline{K} ,\eta \eta,\ldots $ transition amplitudes. 
  
As mentioned, hadronic matrix elements of timelike separated currents, which we refer to as \emph{long-range processes} are also necessary to examine the inner structure of excited QCD states.
For example, the elusive glueballs, hypothesized states composed of pure glue, have been studied in quenched lattice QCD calculations~\cite{Bali:1993fb,Morningstar:1997ff,Morningstar:1999rf} and the  lowest-lying candidate is expected to lie in the $0^{++}$ channel. When the theory is unquenched, these states become hadronic resonances that couple strongly to $\pi\pi$, $K\overline{K}$, $\ldots$ asymptotic states, obscuring any experimental \emph{smoking-gun} evidence of a glueball.
A quantitative measure of the internal charge distribution, which may in turn provide a likelihood of a glueball assignment of a given state, can be extracted from the two-photon coupling.
Given the resonant nature of these states, this coupling needs to be accessed from the same previously mentioned amplitudes for $\gamma^\star \gamma^\star \to \pi\pi,\ldots $, which has been done for the lowest lying scalar resonance, see for instance Refs.~\cite{Pennington:2008xd,Mennessier:2008kk,Dai:2014zta} and Ref.~\cite{Pelaez:2015qba} for a recent review on the extraction of this coupling. 
Another example of a state whose internal structure may be constrained by long-range processes is that of the lowest-lying baryonic resonance, the $\Delta(1232)$. Despite it being an experimentally well studied state, its internal structure is phenomenologically largely unconstrained. A recent proposal was made to access the elastic electromagnetic form factors of this state via the two-photon exchange present in the $e p\to e p\pi$ cross section~\cite{Carlson:2017lys, Machavariani:1999fr, Drechsel:2000um, Drechsel:2001qu, Chiang:2004pw, Pascalutsa:2007wb, Kotulla:2002cg, Pascalutsa:2006up}.
The hadronic contribution can be written in terms of virtual photons, and included in this is the desired $\gamma^\star p\to \Delta \to \gamma^\star p\pi$ resonant amplitude. It is this piece from which one can, in principle, determine the elastic form factors of the $\Delta(1232)$, and subsequently its charge distribution.   
  
These examples, $\gamma^\star \gamma^\star \to \pi\pi,\ldots$ and $\gamma^\star p\to  \gamma^\star p\pi$, fall under a broad class of reactions that can be generically written as $\Jc\to 2+\Jc$ and $1+\Jc\to2+\Jc$ respectively, where $\Jc$ is an external local current, and the 1 and 2 represent the number of hadrons in the initial and final state respectively. In fact, the $\Jc\to 2+\Jc$ reaction can be understood as a simplified case of $1+\Jc\to2+\Jc$, where the initial hadron is replaced with the vacuum. In this work, we present a non-perturbative derivation of the analytic structure of this class of amplitudes. In doing so, we provide an exact closed form for the amplitudes in terms of singular functions that may be determined from the physical subprocesses, together with a priori unknown, smooth, real-valued functions. The results hold for generic systems that may support bound states, resonances, or neither. The derivations follow the formalism presented in Refs.~\cite{Briceno:2019opb, Briceno:2020vgp} for studying the simpler $1+\Jc\to1+\Jc$ and $2+\Jc\to2$ amplitudes which we review. We collectively refer to the classes of amplitudes involving two currents as \emph{Compton-like} amplitudes which we label with the symbol $\Tc$. Although the results presented are indeed exact, they hold for kinematics where only one- and two-body intermediate states may go on-shell, and we only consider hadronic states with zero intrinsic spin.  

The results of the $1+\Jc\to2+\Jc$ amplitude have two immediate applications. First, this serves as a necessary step towards the determination of these amplitudes directly from QCD using lattice QCD. Second, these expressions will provide constraints on the allowed parameterizations of experimental analysis of these reactions. We elaborate further on the first application, since this is expected to be more immediately relevant.
 
Lattice QCD allows for a statistical determination of energies and matrix elements defined in a finite-Euclidean spacetime. Since the physical amplitudes of interest exist in an infinite-Minkowski volume, a framework that connects the lattice QCD calculated matrix elements to these infinite-volume amplitudes is required. Among the classes of amplitudes that are known to be accessible via lattice QCD  are purely hadronic two-~\cite{Luscher:1986pf,Rummukainen:1995vs,Kim:2005gf,Fu:2011xz,He:2005ey,Lage:2009zv,Bernard:2010fp,Briceno:2012yi,Hansen:2012tf,Feng:2004ua,Gockeler:2012yj,Briceno:2014oea,Morningstar:2017spu,2012PhRvD..85k4507L} and three-body scattering amplitudes~\cite{Hansen:2015zga,Mai:2017bge,Briceno:2012rv,Hammer:2017kms}, as well as $1+\Jc\to2$~\cite{Lellouch:2000pv,Briceno:2014uqa,Briceno:2015csa, Briceno:2021xlc} and $2+\Jc\to2$~\cite{Briceno:2015tza,Baroni:2018iau} transition amplitudes. This has already allowed for numerous lattice QCD calculations of resonant systems~\cite{Dudek:2014qha,Guo:2018zss, Silvi:2021uya,Rendon:2020rtw,Alexandrou:2017mpi,Prelovsek:2020eiw,Andersen:2018mau,Brett:2018jqw,Andersen:2017una, Wilson:2014cna,Briceno:2016mjc, Wilson:2019wfr,Wilson:2015dqa, Briceno:2017qmb, Gayer:2021xzv, Dudek:2016cru, Woss:2019hse, Moir:2016srx, Woss:2020ayi, Briceno:2016kkp, Briceno:2015dca, Alexandrou:2018jbt, Niehus:2021iin}. Given the successes of this program (see Refs.~\cite{Briceno:2017max, Hansen:2019nir} for recent reviews), groups have recently begun to consider prospects for studying two-current processes for kinematics where an intermediate two-particle state may go on-shell~\cite{Christ:2015pwa, Feng:2020nqj,Briceno:2019opb,Davoudi:2020xdv,Davoudi:2021noh, Davoudi:2020gxs}. Although these formalisms have not yet been implemented, it is clear that as a preliminary step it will be necessary to have parameterizations of these amplitudes, as well as the amplitudes of the physical subprocesses.  

The remainder of this work is laid out as follows: in Sec.~\ref{sec:main} we present our main results, the on-shell representations for the $1+\Jc\to1+\Jc$ and $1+\Jc\to2+\Jc$ Compton-like amplitudes along with a discussion of the singularity structures that appear in both amplitudes. Moving on to Sec.~\ref{sec:ConsProp} we explore constraints and properties of these on-shell forms, including their analytic continuations, definition of resonance form factors, and the implications of the Ward-Takahashi identity. In Sec.~\ref{sec:Der_Onshell} we present the derivation of our results. Finally, in Sec.~\ref{sec:conclusion} we provide an outlook for these studies.

%%%%%%%%%%%%%%%%%%%%%%%%%%%%%%%%%%%%%%%%%%%
%  Section: Main result
%%%%%%%%%%%%%%%%%%%%%%%%%%%%%%%%%%%%%%%%%%%
\section{Main result}
\label{sec:main}

\begin{figure}[t]
\begin{center}
\includegraphics[width=.9\textwidth]{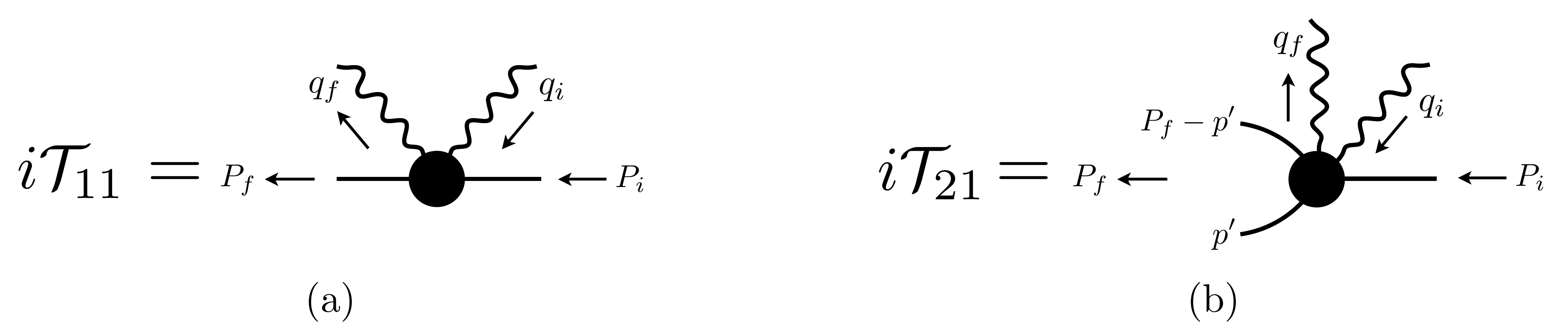}
\caption{ 
Diagrammatic representations of the (a) $1+\Jc\to 1+\Jc$ and (b) $1+\Jc\to 2+\Jc$ amplitudes. The solid lines represent stable single-particle states. The wiggly lines denote external currents. Momentum conservation requires $q_i = P_f + q_f - P_i$.
}
\label{fig:ComptonAmps}
\end{center}
\end{figure}

In this section we present the analytic forms of the amplitudes under consideration where we have singled out the non-analytic pieces that stem from placing intermediate states on their mass shell. We refer to these as their ``on-shell'' forms and save their derivation for Sec.~\ref{sec:Der_Onshell}.
While we assume no specific Lorentz structure for the currents, these amplitudes do depend on the quantum numbers of the currents which we label as $A$ and $B$. These labels, which may or may not be the same, include possible Lorentz indexes as well as other quantum numbers, e.g.\ isospin. We introduce two subscripts for $\Tc$ which label the number of particles in the final/initial state respectively. Thus, using this notation, we will be considering the amplitudes $\Tc_{11}$ and $\Tc_{21}$ which are shown in Fig.~\ref{fig:ComptonAmps} (a) and (b) respectively.

\begin{figure}[t]
\begin{center}
\includegraphics[width=.7\textwidth]{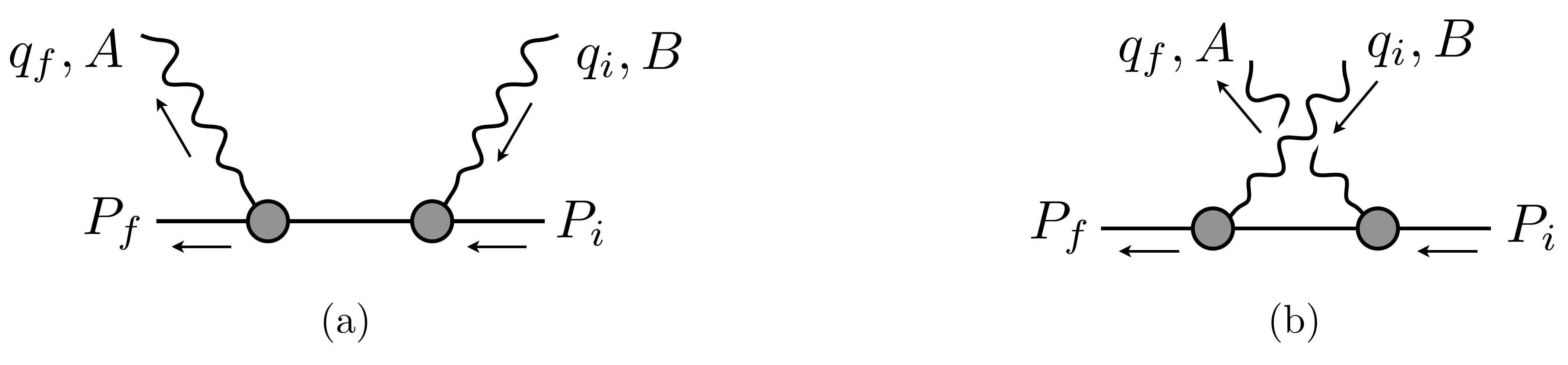}
\caption{ 
Examples of diagrams that appear in the (a) direct and (b) exchange contributions to the $\Tc_{11}$ amplitude.
}
\label{fig:Direct_Exchange}
\end{center}
\end{figure}

We can define the amplitude $\Tc$ in terms of appropriately constructed Fourier transforms of two-current matrix elements between asymptotic states, 
\begin{align}
\label{eq:Tnn_def}
\Tc_{n1}(P_f,P_i,q_f) \equiv i \int\! \diff^4 x \, e^{iq_f \cdot x } \bra{n,P_f;\rm{out}} \mathrm{T} \{ \mathcal{J}^{A} (x) \mathcal{J}^{B} (0) \} \ket{1,P_i;\rm{in}}_{\rm conn.},
\end{align}
where $n$ is either 1 or 2, depending on the number of hadrons in the final state, and ``$\mathrm{T}$" is the time-ordering operator. From the definition, we see that the momentum of the current $\Jc^A$, denoted by $q_f$ here, is leaving the system, while the current $\Jc^B$ injects momentum $q_i = P_f + q_f - P_i$ into the system. Diagrammatically, there are two closely related topologies that contribute to this matrix element, the $s$-channel and the $u$-channel diagrams. We will refer to them as the \emph{direct} and the \emph{exchange} contributions. An example of each of these, for the case of an intermediate single particle state in $\Tc_{11}$, is illustrated in Fig.~\ref{fig:Direct_Exchange} (a) and (b) respectively.

To arrive at the on-shell forms for these amplitudes, we use all-orders perturbation theory in which we assume some generic effective field theory where our hadronic states are stable against strong decay as was done in Ref.~\cite{Briceno:2020vgp}.
We also make two simplifying assumptions here, the first being that our incoming and outgoing hadronic states are spinless, i.e. they can be either scalars or pseudoscalars, and the second being that we are in a kinematic region where intermediate three-particle on-shell states are forbidden. This implies $s=(P_{i}+q_{i})^{2}$ and $u=(P_{i}-q_{f})^{2}$ both lie below the lowest three particle threshold with the appropriate quantum numbers. 
We also only consider spacelike virtualities of the currents, or in the timelike region below any particle production thresholds. 
Since this constraint applies to all of the expressions below we may not make it explicit each time.

\begin{figure}[t]
\begin{center}
\includegraphics[width=.9\textwidth]{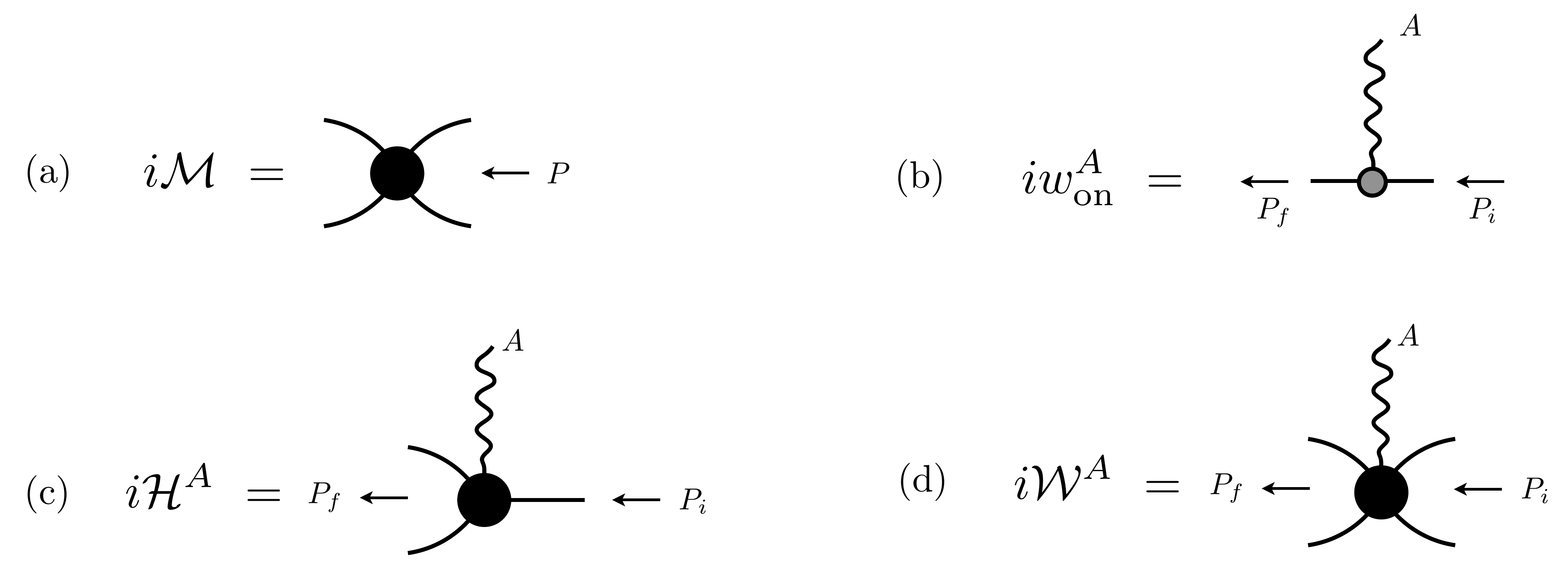}
\caption{ 
Diagrammatic representations of the amplitudes that will appear as building blocks for the Compton-like amplitudes of interest. In (a) we show the $2\to2$ scattering amplitude while the remaining diagrams represent hadrons interacting with external currents. In increasing order of complexity these are the (b) $1+\Jc\to 1$, (c) $1+\Jc\to 2$, and (d) $2+\Jc\to 2$ amplitudes.
}
\label{fig:BuildingBlocks}
\end{center}
\end{figure}

The final expressions for the Compton-like amplitudes depend on the amplitudes describing the kinematically allowed sub-processes. These are the purely hadronic $2\to2$ amplitudes, and the $1+\Jc\to 1$, $1+\Jc\to 2$ and $2+\Jc\to 2$ transition amplitudes involving a single current insertion, illustrated in Fig.~\ref{fig:BuildingBlocks}. These amplitudes, which we respectively label as $\mathcal{M}$, $w_{\mathrm{on}}$, $\mathcal{H}$, and $\mathcal{W}$, were the focus of Ref.~\cite{Briceno:2020vgp}.
As discussed in detail in the aforementioned reference, $\mathcal{W}$ has simple pole singularities that can be expressed in terms of $w_{\mathrm{on}}$, $\Mc$, and single-particle propagators. The remainder of the amplitude is denoted by $\Wc_\df$, where the subscript stands for ``divergence free".

In Sec.~\ref{sec:Der_Onshell} we provide integral equations for $\Mc$ and $\Hc$, since they play an important role through the rest of the derivation. For completeness, here we provide the on-shell expressions for each of the amplitudes depicted in Fig.~\ref{fig:BuildingBlocks},
\begin{align}
\label{eq:M_on-shell}
i\Mc(s)
&=
i\Kc(s) \,
\frac{1}{1 - i \rho \, \Kc(s)} \, ,
%%%%%%%%%%%%%%%%%%%%%%%%%%
\\
\label{eq:w_on}
w_{\mathrm{on}}^{A}(P_{f},P_{i})
&=
\sum_{j} K_{j}^{A}(P_{f},P_{i})f_{j}(Q^{2}),
%%%%%%%%%%%%%%%%%%%%%%%%%%
\\
\label{eq:H_on-shell}
i\Hc^{A} (P_f,P_i)
&=
i\Mc (s_f) \mathcal{A}^{A}_{21}  (P_f,P_i) \, ,
%%%%%%%%%%%%%%%%%%%%%%%%%%
\\
\label{eq:Wdf_def}
i\Wc_{\df}^{A}(P_f,P_i) 
&= 
\Mc (s_f) \, \left[ \,  i\Ac_{22}^{A}(P_f,P_i) 
+ \sum_{j} if_{j}(Q^2) \Gc_{j}^{A}(P_f,P_i)  \, \right] \, \Mc(s_i) \, ,
\end{align}
where $\Kc$ is the two-body $K$-matrix, and $\Ac_{21}$ and $\Ac_{22}$ are the single-current analogs of the $K$-matrix where the subscripts indicate how many hadrons are in the final and initial state respectively. Each of these objects are real and smooth functions in the kinematic domain of interest, but in principle contain singularities away from this region arising from crossed channel processes or higher multi-particle thresholds.~\footnote{The $K$-matrix can have unphysical simple poles in this kinematic region, but as discussed in the derivation, the $\Ac$ functions cannot have such poles.}.
Finally, $K_{j}$ are kinematic functions whose Lorentz structure depends on that of the current, and $f_{j}$ are the single particle form factors which depend on $Q^{2}=-(P_{f}-P_{i})^{2}$.
For a given Lorentz structure, the decomposition of $w_{\rm on}$ will contain a finite set of linearly independent $K_j$ tensors which we enumerate with the subscript $j$.
We also use the conventional notation $s_f\equiv P_f^2$ and $s_i\equiv P_i^2$.
In these expressions the two-particle states have been partial-wave projected and the amplitudes are matrices or vectors in angular momentum space accordingly. The label ``on'' in $w_{\rm on}$ emphasizes that it has been projected on-shell such that the form factors $f_j$ only depend on the virtuality of the current, but the kinematic function depends on the momenta $P_{f/i}$ even when it is off-shell. 

The singularities of these functions are encoded in the two-particle phase space factor $\rho$, and the triangle function~$\Gc$. When only one two-particle channel is kinematically allowed to go on its mass shell, these can be written as~\footnote{For a general expression and deeper discussion of the singularities we point the reader to Ref.~\cite{Briceno:2020vgp}}
\begin{align}
\rho_{\ell' m_{\ell'};\ell m_{\ell}} 
&=
\delta_{\ell'\ell}\, \delta_{m_{\ell'}m_{\ell}}
\frac{\xi \, q^{\star}}{8\pi \sqrt{s}}  ,
\label{eq:ps}
%%%%%%%%%%%%%%%%%
\\
    \label{eq:G.function}
\Gc_{j,2; \ell' m_{\ell'} ; \ell m_{\ell}}^{A} (P_f,P_i) 
&\equiv 
\int\! \frac{\diff^4k}{(2\pi)^4}  
 \,  \frac{ \mathcal Y_{\ell' m_{\ell'}}^{*}({\mathbf{k}}_f ^{\star} )  \,\,\, iK_{j,2}^{A}( k_{f},k_{i})  \,\,\, \mathcal{Y}_{\ell m_{\ell}}({\mathbf{k}}_i^{\star} ) }{(k^2 - m_1^2 + i\epsilon)(k_f^2 - m_2^2 + i\epsilon) (k_i^2 - m_2^2 + i\epsilon)} \, ,
\end{align}
where $q^\star$ is the two-particle relative momentum in the center-of-momentum (CM) frame, the symmetry factor $\xi$ is defined to be $1/2$ if the particles in this channel are identical and $1$ otherwise, and $\Yc_{\ell,m_{\ell}}$ are proportional to the solid harmonics, as introduced in Refs.~\cite{Briceno:2015tza, Baroni:2018iau, Briceno:2020vgp}
\begin{equation}
\label{eq:sph_w_barrier}
\Yc_{\ell m_{\ell}}({\mathbf{k}} ^{\star} )
=
\sqrt{4\pi} \, Y_{\ell m_{\ell}}(\hat{\mathbf{k}} ^{\star} )
 \, \left(  \frac{k^{\star}}{q^{\star}} \right)^{\ell} \,,
\end{equation}
where the centrifugal barrier factors remove the spurious threshold singularities of the spherical harmonics.
As shown explicitly in Ref.~\cite{Briceno:2020vgp}, the $\Gc$ function encodes a logarithmic singularity.  
The last thing to point out here is that the numbered subscripts that appear in Eq.~\eqref{eq:G.function} are used to distinguish between the particles with masses $m_{1}$ and $m_{2}$ in the intermediate state, with the number appearing on $\Gc$ and $K_j$ being the particle that the current is coupling to. Thus Eq.~\eqref{eq:G.function} is specific to the case where the current couples to particle 2.

Having reviewed the amplitudes and components that will appear as building blocks, we now move on to the main focus of this work. Starting with the simplest Compton amplitude, $\Tc_{11}$, in Sec.~\ref{sec:1Jto1J} we re-derive the expression obtained in Ref.~\cite{Briceno:2019opb},
\begin{multline}
    \label{eq:T11_on-shell}
    i\Tc_{11}(P_{f},P_{i},q_{f})
    =
    i\Bc_{11} (P_{f},P_{i},q_{f})
    +
    iw_{\mathrm{on}}^{A}(P_{f},P_{s})iD(s)iw_{\mathrm{on}}^{B}(P_{s},P_{i})
        +
    iw_{\mathrm{on}}^{B}(P_{f},P_{u})iD(u)iw_{\mathrm{on}}^{A}(P_{u},P_{i})\\
    \qquad+
    \Ac_{12}^{A}(P_{f},P_{s})i\Mc(s)\Ac_{21}^{B}(P_{s},P_{i})
    +
    \Ac_{12}^{B}(P_{f},P_{u})i\Mc(u)\Ac_{21}^{B}(P_{u},P_{i})
    ,
\end{multline}
where $D$ is the simple pole contribution of the single-particle propagator for a particle with mass $m$,
\begin{align}
    \label{eq:D_prop}
    iD(k^{2})
    &=
    \frac{i}{k^{2}-m^{2}+i\epsilon},
\end{align}
and $\Bc_{11}$ is a real-valued function~\footnote{In Ref.~\cite{Briceno:2019opb}, this function was labeled $\mathbf{S}$.} whose singularities lie outside of the kinematic region considered, similar to $\Kc$, $\Ac_{21}$, and $\Ac_{22}$. As with the $1+\Jc\to 1$ amplitude, one can perform a Lorentz decomposition of $\Bc_{11}$ to write this in terms of a sum over products of Lorentz tensors and generalized form factors. We have also introduced the notation $P_{s}=P_f+q_f=P_i+q_i$ and $P_{u}=P_{f}-q_{i}=P_{i}-q_{f}$ such that $P_{s}^{2}=s$ and $P_{u}^{2}=u$.

From Eq.~\eqref{eq:T11_on-shell} we can see that there are two sources of singularities that occur in the $\Tc_{11}$ amplitude.
The second and third terms indicate the possibility of a simple pole singularity originating from the pole piece of the propagator.
The last two terms can also contribute singularities of similar structure if $\Mc$ features bound state poles, but $\Mc$ also contains branch points corresponding to two-particle thresholds. \footnote{In this expression the single intermediate state is assumed to have quantum numbers different than the two-particle intermediate state, otherwise this contribution will be double counted by the bound state pole in $\Mc$. An example of a reaction with these two different contributions is the Compton scattering off a pion. In that case an intermediate pion cannot mix with a two pion state.}

Moving on to the main new result of this work, we show in Sec.~\ref{sec:1Jto2J} that the on-shell expression for the Compton-like amplitude $\Tc_{21}$ can be written as,
\begin{multline}
    \label{eq:T21_on-shell}
    i\Tc_{21}(P_f,\hat{\vec{p}}^{\prime\star}_f;P_i;q_f)
   = 
   i\Hc^{A}_{\rm on}(P_{f},\hat{\vec{p}}^{\prime\star}_f;P_{s}) iD({s}) iw^{B}_{\mathrm{on}}(P_{s},P_{i})
   +
   i\Hc^{B}_{\rm on}(P_{f},\hat{\vec{p}}^{\prime\star}_f;P_{u}) iD({u}) iw^{A}_{\mathrm{on}}(P_{u},P_{i})
    \\
    +
   \sum 
   \left\{
   iw_{\mathrm{on}}iDi\overline{\Hc}
   \right\}
    +i\Tc_{21,\df} (P_f,\hat{\vec{p}}^{\prime\star}_f;P_i;q_f)
    \,,
\end{multline}
where we have introduced the notation $p'_{x}\equiv P_{x}-p'$ where $x$ can be $f$, $s$, or $u$. We use semicolons to distinguish between the dependence on the final state, initial state, and the current, since at least one of them depends on multiple momenta.
In the first two terms we introduce $\Hc_{\rm on}$ which is the extension of $\Hc$ to off-shell values of the momentum describing the single particle state where, similarly to $w_{\rm on}$ defined in Eq.~\eqref{eq:w_on}, the energy-dependent form factors are kept on-shell. A thorough discussion of this on-shell projected amplitude is given in appendix~\ref{sec:App.osproj}.
The sum in the third term indicates that we need to sum over the current coupling to each of the two external legs in the final state, both for direct and exchange contributions. Therefore in the case where the current only couples to the external particle with final momentum $p'_{f}$ this term should look like,
\begin{figure}[t]
\begin{center}
\includegraphics[width=.83\textwidth]{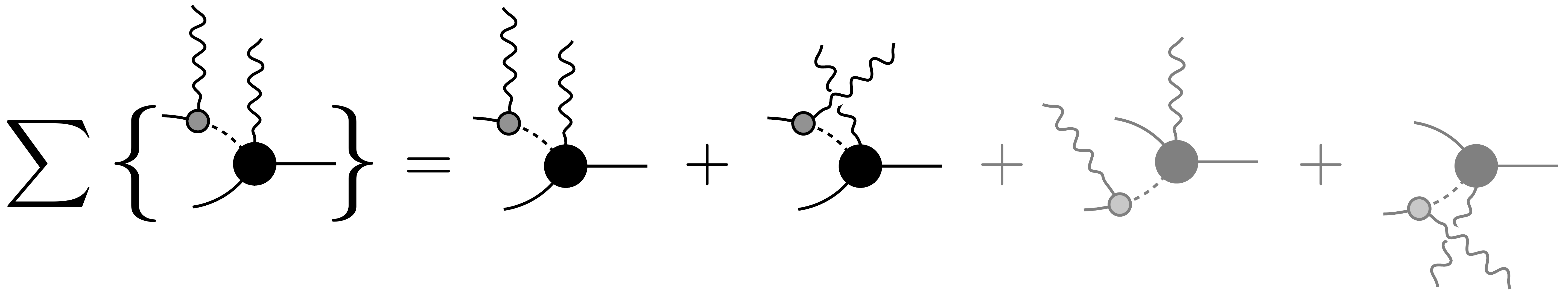}
\caption{ 
Diagrammatic representation of the expansion of the sum shown on the left-hand side of Eq.~\eqref{eq:wDH} where one of the external currents couples to a single particle in the final state. Shown are all four of the possible terms contained within the sum; however, the two final terms are faded as they are not included in Eq.~\eqref{eq:wDH} since we make the assumption that only a single final state particle is charged. The grey circle represents the amplitude $w_{\mathrm{on}}$, the dotted line represents the pole piece of the single particle propagator $D$, and the dark circles are the $1+\Jc\to2$ transition amplitude but with additional barrier factors as designated by the bar in $\overline{\mathcal{H}}$. 
}
\label{fig:wDH}
\end{center}
\end{figure}
\begin{equation}
\label{eq:wDH}
       \sum 
   \left\{
   iw_{\mathrm{on}}iDi\overline{\Hc}
   \right\}
   = iw^A_{\mathrm{on}}(p'_f,p'_s) iD(p'^2_s) i\overline{\Hc}^B(P_s,{\vec{p}}^{\prime\star}_s;P_i)
     + iw^B_{\mathrm{on}}(p'_f,p'_u) iD(p'^2_u) i\overline{\Hc}^A(P_u,{\vec{p}}^{\prime\star}_u;P_i),
\end{equation}
where $\overline{\Hc}$ is the $1+\Jc\to 2$ transition amplitude with additional barrier factors in its partial wave projection to cancel out spurious threshold singularities arising from the spherical harmonics. This is shown diagrammatically in Fig.~\ref{fig:wDH}. The precise definition is given in Sec.~\ref{sec:T211B} in Eq.~\eqref{eq:Hoverldef}.

Finally, we have labeled the last term in Eq.~\eqref{eq:T21_on-shell} with the subscript ``df" which, as previously mentioned,  stands for ``divergence free''. The partial wave projection of this term, as shown in Sec.~\ref{sec:1Jto2J}, can be written as,
\begin{align}
    \label{eq:T21_df}
    i\Tc_{21,\df}(P_f,P_i,q_f)
    &= 
    i\Mc(s_f)\Bc_{21}(P_f,P_i,q_f)
    % &\hspace{-1.5cm}
    +
    i\Wc^A_{\df}(P_f,P_s)\Ac^B_{21}(P_s,P_i)\,
    +
    i\Wc^B_{\df}(P_f,P_u)\Ac^A_{21}(P_u,P_i)\,,
\end{align}
where $\Bc_{21}$ is a new, smooth, real-valued function which depends on the total momentum of both the initial and final states as well as the momentum and Lorentz structure of both of the currents. 

Looking at Eqs.~\eqref{eq:T21_on-shell} and~\eqref{eq:T21_df} we can see that $\Tc_{21}$ inherits its singularity structure from the previously presented sub-amplitudes. The first three terms in Eq.~\eqref{eq:T21_on-shell} correspond to one of the currents coupling to one of the external legs. In each of these cases we get a singularity from the pole piece of the single-particle propagator, $D$. The other singularities for this amplitude reside in $\Tc_{21,\df}$. Both $\Mc$ and each of the $\Wc_{\df}$'s will have threshold singularities in $s_{f}$; however, the $\Wc_{\df}$'s will also have threshold singularities in $s$ and $u$ respectively as well as the logarithmic singularities contained in the triangle function.

Our choice for the on-shell projected amplitudes of the sub-processes $w_{\rm on}$ and $\Hc_{\rm on}$ is not unique, especially when the amplitudes obey a constraint like gauge invariance. This freedom however does not modify the location, strength, and nature of the singularities that appear in our main results. Different prescriptions simply change how smooth contributions are shared between different terms of the on-shell projection. Our prescription choice, and possible alternatives, are described in App.~\ref{sec:App.osproj}.

The remainder of this work focuses on discussing properties of these Compton-like amplitudes as well as deriving the on-shell forms given in Eqs.~\eqref{eq:T11_on-shell}, ~\eqref{eq:T21_on-shell}, and~\eqref{eq:T21_df}.

%%%%%%%%%%%%%%%%%%%%%%%%%%%%%%%%%%%%%%%%%%%
%  Section: Constraints and Properties
%%%%%%%%%%%%%%%%%%%%%%%%%%%%%%%%%%%%%%%%%%%
\section{Constraints and Properties}
\label{sec:ConsProp}

Here we present further analysis of the expressions for the Compton-like amplitudes given in the previous section. First we discuss the analytic continuation of the amplitudes, this is required for studying the properties of the dynamical resonances featured within an amplitude. We also use this to show that the formalism presented here is consistent with previous work. Finally, we discuss the Ward-Takahashi identity as it relates to these amplitudes when considering conserved vector currents and the additional constraints it presents.

%%%%%%%%%%%%%%%%%%%%%%%%%%%%%%%%%%%%%%%%%%%
%  Subsection: Analytic Continuation
%%%%%%%%%%%%%%%%%%%%%%%%%%%%%%%%%%%%%%%%%%%
\subsection{Analytic Continuation}
\label{sec:AnalyticContinuation}

\begin{figure}[t]
\begin{center}
\includegraphics[width=.9\textwidth]{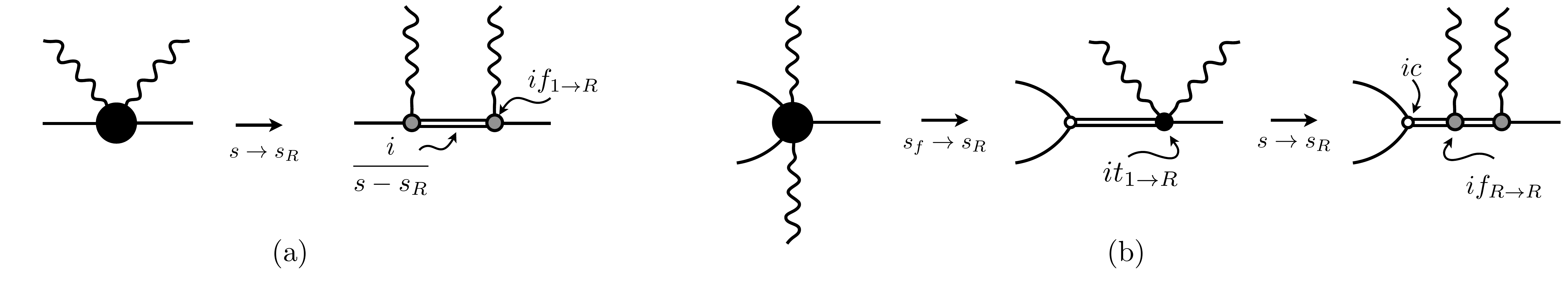}
\caption{ 
Diagrammatic representations of the (a) $1+\Jc\to 1+\Jc$ and (b) $1+\Jc\to 2+\Jc$ amplitudes when approaching a resonance pole. The double lines represent the propagator for the resonance state, the black circles represent two-current scattering amplitudes, the gray circles represent single-current form factors, and the open circle is the purely hadronic coupling between the resonance and the asymptotic two-particle state. 
}
\label{fig:resonances}
\end{center}
\end{figure}

In this section we present the analytic continuation of the amplitudes in the case they contain a resonant intermediate state. For simplicity we show this for the case of a scalar current with an $S$-wave resonance; however, the steps shown are valid for currents of any Lorentz structure as well as systems in higher partial waves.

Starting with the forward limit of the $1+\Jc\to1+\Jc$ amplitude, i.e.\ $q_{i}=q_{f}$ and defining $Q^2=-q_i^2=-q_f^2$, as one approaches the resonance pole we find,
\begin{align}
	\label{eq:lim_T11}
	\lim_{s\to s_{R}}(s-s_{R}) \Tc_{11}^{\rm II}(s,Q^{2})
	=
	{-[f_{1\to R}(Q^{2})]^{2}},
\end{align}
where $\Tc_{11}^{\rm II}$ is the analytic continuation of $\Tc_{11}$ to the second Riemann sheet in $s$, $f_{1\to R}(Q^{2})$ is the transition form factor, and $s_{R}$ is the location of the resonance pole.
This is also shown diagrammatically in Fig.~\ref{fig:resonances}(a) where the double line represents the resonance propagator and each vertex is equal to the transition form factor.
The analytic continuation of $\Tc_{11}$ is obtained from the knowledge of its analytic structure given by its on-shell representation.
By inspecting Eq.~\eqref{eq:T11_on-shell} one can notice that taking the analytic continuation of the amplitudes $\Mc$ therein is sufficient to obtain $\Tc^{\rm II}_{11}$.
We solve for the transition form factor in Eq.~\eqref{eq:lim_T11} and
exploit that the behavior of the scattering amplitude $\Mc$ close to the resonance is
\begin{align}
    \lim_{s\to s_{R}} (s-s_{R}) \Mc^{\rm II}(s)
    =
    -c^2,
\end{align}
where $\Mc^{\rm II}$ is the analytic continuation of $\Mc$ to the second sheet in $s$, and $c$ is the coupling between the resonance and the external two-particle state. Only the term featuring the amplitude $\Mc^{\rm II}(s)$ in the continuation of Eq.~\eqref{eq:T11_on-shell} will survive the limit, to find
%Solving this for the transition form factor gives,
\begin{align}
	\label{eq:trans_ff}
	f_{1\to R}(Q^{2})
    =
	c\Ac_{21}(s_{R},Q^{2}).
\end{align}
Equation~\eqref{eq:trans_ff} agrees with Eq.~(20) of Ref.~\cite{Briceno:2020vgp} where $f_{1\to R}$ was found from the properties of the $\Hc$ amplitude close to the resonance, thus providing a consistency check for Eq.~\eqref{eq:T11_on-shell}.

Moving on to the $1+\Jc\to2+\Jc$ amplitude we find that a resonance can couple to the final state as well as to intermediate states. This implies that the amplitude contains a resonance pole both in the final two-state energy squared $s_f$, as well as in the intermediate energy squared $s$.\footnote{A resonance could also appear in the exchange channel as a pole in variable $u$, for simplicity here we restrict ourselves to the production in the direct diagrams.}
We will study each of these poles one at a time by taking the limits to the resonance in two steps
\begin{align}
    \label{eq:lim_T21_sf}
    \lim_{s_{f}\to s_{R}}
    (s_{f}-s_{R})\Tc_{21}^{\rm II,I}(P_{f},P_{i},q_{f})
    &=
    -c\,{t}_{1\to R}(Q_{f}^{2},Q_{i}^{2},s),\\
\lim_{s\to s_{R}}
    \label{eq:lim_T21_s}
    (s-s_{R})t^{\rm II}_{1\to R}(Q_{f}^{2},Q_{i}^{2},s)
    &=
    -f_{R\to R}(Q_f^{2})
    f_{1\to R}(Q_i^{2}),
\end{align}
where $Q_{f/i}^{2}=-q_{f/i}^{2}$, and $\Tc_{21}^{\rm II,I}$ is the analytic continuation of $\Tc_{21}$ to the second Riemann sheet in the $s_{f}$ variable only, ${t}_{1\to R}$ is the Compton-like amplitude coupling the single-particle state to the resonance, and ${t}^{\,\rm II}_{1\to R}$ is its analytic continuation to the second sheet of variable $s$. This Compton-like transition is a new quantity that has not previously been considered. As previously mentioned, this can have dynamical singularities as well.
We also find that the second limit allows for access to the same elastic resonant form factors, $f_{R\to R}$, that can be obtained from $\Wc$. The diagrammatic representation of the amplitudes close to these limits is shown in Fig.~\ref{fig:resonances}(b). 
The Compton-like transition can be found by solving Eq.~\eqref{eq:lim_T21_sf},
\begin{align}
	\label{eq:new_t1R}
    t_{1\to R}(Q_{f}^{2},Q_{i}^{2},s)
    &=
    \lim_{s_{f}\to s_{R}}
    \frac{s_{R}-s_{f}}{c}\Tc_{21}^{\rm II,I}(P_{f},P_{i},q_{f}) \nn\\
    &=
    c\Ac_{21}^{A}(s_{R},Q_{f}^{2})iD(s)iw_{\rm on}^{B}(Q_{i}^{2})
    +
    c\Ac_{21}^{B}(s_{R},Q_{i}^{2})iD(u)iw_{\rm on}^{A}(Q_{f}^{2})
    \nn\\
    &\qquad\qquad+
    \lim_{s_{f}\to s_{R}}
    \frac{s_{R}-s_{f}}{c}
    \Tc_{21,\df}^{\rm II,I}(P_{f},P_{i},q_{f}).
\end{align}
The final term can be written explicitly as,
\begin{align}
    \label{eq:lim_T21_II,I}
    \lim_{s_{f}\to s_{R}}
    \frac{(s_{R}-s_{f})}{c}
    \Tc_{21,\df}^{\rm II,I}(P_{f},P_{i},q_{f})
    &=
    \lim_{s_{f}\to s_{R}}c\Bc_{21}(P_{f},P_{i},q_{f})
    +
    \frac{(s_{R}-s_{f})}{c}
    \Wc_{\df}^{A,{\rm II,I}}(s_{f},Q_{f}^{2},s)\Ac_{21}^{B}(s,Q_{i}^{2})\nn\\
    &\qquad\qquad+
    \frac{(s_{R}-s_{f})}{c}
    \Wc_{\df}^{B,{\rm II,I}}(s_{f},Q_{i}^{2},u)\Ac_{21}^{B}(u,Q_{f}^{2})\nn\\
    &=
    c(\Bc_{21}(Q_{f}^{2},Q_{i}^{2},s)
    +
    [\Ac_{22}^{A}(s_{R},Q^{2}_f,s)+f(Q_{f}^{2})\Gc^{A, \rm II,I}(s_{R},Q_{f}^{2},s)]\Mc(s)\Ac_{21}^{B}(s,Q_{i}^{2})\nn\\
    &\qquad\qquad+
    [\Ac_{22}^{B}(s_{R},Q^{2}_i,u)+f(Q_{i}^{2})\Gc^{B, \rm II,I}(s_{R},Q_{i}^{2},u)]\Mc(u)\Ac_{21}^{A}(u,Q_{f}^{2})).
\end{align}
Here again $\Wc_{\df}^{\rm II,I}$ and $\Gc^{\rm II,I}$ are the analytic continuations of these functions to the second Riemann sheet, but in the $s_{f}$ variable only. As is shown explicitly in Appendix~\ref{sec:App.AC}, $\Gc^{\rm II,I}$ is given by 
\begin{align}
    \Gc^{\rm II,I}(s_{f},Q_{f}^{2},s)
    &=
    \Gc(s_{f},Q_{f}^{2},s)
    -
    {\rm Disc}_{s_{f}} \Gc(s_{f},Q_{f}^{2},s),
\end{align}
where ${\rm Disc}_{s_{f}}$ is the discontinuity across the branch cut on the $s_{f}$-axis.
Combining Eqs.~\eqref{eq:new_t1R} and~\eqref{eq:lim_T21_II,I} we find 
\begin{multline}
    \label{eq:t_1R}
    t_{1\to R}(Q_{f}^{2},Q_{i}^{2},s)
    =
    c\left[
    \Bc_{21}(Q_{f}^{2},Q_{i}^{2},s)
    +
    \Ac_{21}^{A}(s_{R},Q_{f}^{2})iD(s)iw_{\rm on}^{B}(Q_{i}^{2})
    +
    \Ac_{21}^{B}(s_{R},Q_{i}^{2})iD(u)iw_{\rm on}^{A}(Q_{f}^{2})\right.\\
    \left.+
    [\Ac_{22}^{A}(s_{R},Q^{2}_f,s)+f(Q_{f}^{2})\Gc^{A, \rm II,I}(s_{R},Q_{f}^{2},s)]\Mc(s)\Ac_{21}^{B}(s,Q_{i}^{2})\right.\\
    \left.+
    [\Ac_{22}^{B}(s_{R},Q^{2}_i,u)+f(Q_{i}^{2})\Gc^{B, \rm II,I}(s_{R},Q_{i}^{2},u)]\Mc(u)\Ac_{21}^{A}(u,Q_{f}^{2}))
    \right].
\end{multline}
In the case where the resonance becomes a stable bound state Eq.~\eqref{eq:t_1R} has to have the same analytic structure as given by \eqref{eq:T11_on-shell}. The first three terms in these equations already meet this requirement, and it can be shown that the last two terms also share the same analytic structure up to an additive smooth contribution, which can be reabsorbed into the $\Bc$ term.

Having an on-shell representation of $t_{1\to R}$, we may now use this result along with Eq.~\eqref{eq:lim_T21_s} to access the elastic resonant form factor, $f_{R\to R}$,
\begin{align}
    f_{R\to R}(Q_{f}^{2})
    &=
    \lim_{s\to s_{R}} (s_{R}-s)
    \frac{t_{1\to R}^{\rm II}(Q_{f}^{2},Q_{i}^{2},s)}{f_{1\to R}(Q_{i}^{2})}\nn\\
    &=
    \frac{c^{3}(\Ac_{22}^{A}(s_{R},Q_{f}^{2},s_{R})+f(Q_{f}^{2})\Gc^{A, \rm II,II}(s_{R},Q_{f}^{2},s_{R}))\Ac_{21}^{B}(s_{R},Q_{i}^{2})}{c\Ac^B_{21}(s_{R},Q_{i}^{2})}\nn\\
    &=
    c^{2}(\Ac_{22}^{A}(s_{R},Q_{f}^{2},s_{R})+f(Q_{f}^{2})\Gc^{A, \rm II,II}(s_{R},Q_{f}^{2},s_{R})).
\end{align}
In the second equality, we only kept the terms in $t_{1\to R}^{\rm II}$ that survive the limit and we used the definition of $f_{1\to R}$ from Eq.~\eqref{eq:trans_ff}. In the last equality, $\Gc^{A, \rm II,II}$ has been analytically continued to the second sheet for both $s$ and $s_f$. It is important to note that the final result agrees with the definition of the $f_{R\to R}$ found in Eq.~(25) of Ref.~\cite{Briceno:2020vgp}, providing further evidence for the expression found in Eq.~\eqref{eq:T21_on-shell}. 

%%%%%%%%%%%%%%%%%%%%%%%%%%%%%%%%%%%%%%%%%%%
%  Subsection: Ward-Takahashi Identity
%%%%%%%%%%%%%%%%%%%%%%%%%%%%%%%%%%%%%%%%%%%
\subsection{Ward-Takahashi Identity}
\label{sec:WTI}

In this section we will discuss the implication of the Ward-Takahashi Identity of conserved vector currents, i.e.\ gauge invariance, to our results. In the case of the initial and final current insertions corresponding to external on-shell photons, we will label the amplitudes as $\Tc^{\mu\nu}_{n1,\mathrm{R}}(P_f,P_i,q_f)$ where the vector index $\mu$ corresponds to the outgoing photon, while the index $\nu$ to the incoming one.
The restrictions imposed by gauge invariance on $1 +\Jc^\nu \to 1 +\Jc^\mu$ are
\begin{equation}
   q_{f,\mu} \Tc_{11}^{\mu\nu} (P_f,P_i,q_f)
   =q_{i,\nu} \Tc_{11}^{\mu\nu} (P_f,P_i,q_f) =0\,,
\end{equation}
which give rise to a series of low-energy theorems on the amplitudes \cite{Low:1954kd,GellMann:1954kc,Fearing:1996gs}, these apply even for off-shell photons. In particular, by a clever choice of the kinematic tensors, the dynamics of real Compton scattering off spinless mesons can be contained within only two scalar amplitudes \cite{Lvov:2001zdg}. These amplitudes are free of kinematic singularities, and satisfy gauge invariance, time inversion, parity and charge conjugation. This is achieved by choosing a set of kinematic tensors that incorporate gauge invariance explicitly. However, the on-shell expansion and the gauge invariance constraints will prove sufficient to recover the forward limit of the amplitude.

As shown explicitly in App.~\ref{sec:App.Loren}, expanding Eq.~\eqref{eq:T11_on-shell} around zero photon energy $q_0$, in the rest frame of the hadron, yields the expression
\begin{equation}\label{eq:WTT11lpe}
    i\Tc^{\mu\nu}_{11,\mathrm {R}}(P,P,q) = 2if(0)^2\qty(g^{\mu\nu}- (P^\mu q^\nu+P^\nu q^\mu)\frac{1}{mq_0}) +\mathcal{O}(q_0^2)\,,
\end{equation}
where the second term in the parenthesis vanishes once this amplitude is contracted with the external photon wavefunctions. The derivation of Eq.~\eqref{eq:WTT11lpe} does not require an explicit Lorentz decomposition of $T_{11,R}^{\mu\nu}$, but only the analytic properties of each of the terms of Eq.~\eqref{eq:T11_on-shell}. This demonstrates that the constraints from the integral equation Eq.~\eqref{eq:1Jto1J.T11_dse} plus gauge invariance reproduce the well-known Thomson scattering
\begin{equation}
\epsilon_\mu(0)^{\prime*}\epsilon_\nu(0) i\Tc^{\mu\nu}_{11,\mathrm {R}}(P,P,0) = 2if(0)^2\vec{\boldsymbol{\epsilon}}^{\prime*}\cdot \vec{ \boldsymbol{\epsilon}}\,,
\end{equation}
where $\boldsymbol{\epsilon}^{(\prime)}$ is the polarization three-vector of the incoming (outgoing) photon.

Similarly, the Ward identity for the $1+\Jc^\nu\to2+\Jc^\mu$ amplitude makes it vanish when contracted against the external photon momentum, regardless of the virtuality of the photon
\begin{equation}
    q_{f,\mu}i\Tc_{21}^{\mu\nu}(P_f,p';P_i;q_f) = 
    q_{i,\nu}i\Tc_{21}^{\mu\nu}(P_f,p';P_i;q_f) = 0\,.
\end{equation}
This means that the divergent free part is equal to the negative of the long-distance contributions, whenever they are contracted with the momenta of the external photons
\begin{multline}\label{eq:WT.qdflg}
    q_{f,\mu} i\Tc^{\mu\nu}_{21,\mathrm{df}}(P_f,p';P_i;q_f) =  -q_{f,\mu}\Big[i\mathcal{H} ^{\mu}_{\rm on}(P_f,p';P_s) iD(s) iw_\mathrm{on}^{\nu}(P_s,P_i) +  iw_\mathrm{on}^{\mu}(p'_f,p'_s)   iD(p^{\prime 2}_s) i\overline{\mathcal{H} }^{\nu}(P_s,p';P_i)\\
    +i\mathcal{H}_\mathrm{on}^{\nu}(P_f,p';P_u) iD(u) iw_\mathrm{on}^{\mu}(P_u,P_i) +  iw_\mathrm{on}^{\nu}(p'_f,p'_u)   iD(p^{\prime2}_u) i\overline{\mathcal{H} }^{\mu}(P_u,p';P_i)
    \Big]\,,
\end{multline}
and similarly when contracted with $q_{i,\nu}$.
Equation~\eqref{eq:WT.qdflg} includes both the direct channel as well as the crossed channel contributions. The barrier factors in $\overline{\Hc}^\mu$, as seen from Eq.~\eqref{eq:T21.Hbar}, are an overall multiplicative factor in each on-shell partial-wave transition amplitude, and as a result do not affect the behavior of the amplitudes under the Ward identity. On the other hand, in Appendix~\ref{sec:App.osproj} we discuss the implications of the on-shell projection of single particle states in $\Hc_{\rm on}$ and $w_{\rm on}$ when contracting the sub-amplitudes against the external photon momentum. For the case of a spinless particle, the Lorentz decomposition of the on-shell projected $1+\Jc^\mu\to 1$ amplitude is
\begin{equation}\label{eq:WT.wonLD}
    w^\mu_\mathrm{on}(k_f,k_i) = (k_f+k_i)^\mu f(-(k_f-k_i)^2)\,.
\end{equation}
This can be used to calculate the contractions between the current momenta and the single-particle currents,
\begin{align}
    q_{f,\mu} w_\mathrm{on}^{\mu}(p'_f,p'_s)
    & = (p'_s-p'_f)_\mu w_\mathrm{on}^{\mu}(p'_f,p'_s) =(p^{\prime 2}_s-m^2)f(Q_f^2)\,,\\
    q_{f,\mu} w_\mathrm{on}^{\mu}(P_u,P_i)
    & = (P_i-P_u)_\mu w_\mathrm{on}^{\mu}(P_u,P_i)
    = (m^2-u)f(Q_f^2)\,,
\end{align}
where the right most equalities follow from the fact that the momenta $P_i$ and $p'_f=P_f-p'$ are on shell.

In the case of spinless particles, the Lorentz decomposition of the $1+\Jc^\mu\to 2$ depends on the intrinsic parity of the hadrons. We will focus on the the case where the three hadrons are pseudoscalars, and leave in App.~\ref{sec:App.osproj} the relevant formulae to derive the case with opposite parity, although the final result for both cases is equal. The case with pseduoscalars is the most relevant in the light quark sector since  all scalar mesons made up of only light quarks are hadronic resonances.
The Lorentz decomposition of this on-shell projected transition is given by
\begin{equation}
    i\Hc^\mu_{\rm on}(P_f,p';P_s) = \epsilon^{\mu\nu\sigma\rho}P_{f,\nu}p'_{\sigma}P_{s,\rho} ih(s_f,(P_s-p')^2,q_f^2),
\end{equation}
where the function $h$ is an energy dependent transition form factor. The contraction of this amplitude with the current momenta $q_f=P_f-P_s$ will vanish due to the Levi-Civita tensor, even for $P_s\neq m^2$.
After performing the momentum contraction to the long range terms in Eq.~\eqref{eq:WT.qdflg}, and the equivalent operation with $q_{i,\nu}$, they simplify to
\begin{align}
q_{f,\mu} i\Tc^{\mu\nu}_\text{df}(P_f,p';P_i;q_f) &= f(Q_f^2)(i\overline{\Hc}^\nu(P_s,p';P_i) - i{\Hc}_\mathrm{on}^\nu(P_f,p';P_u))\,,\label{eq:qfTsimp}\\
q_{i,\nu} i\Tc^{\mu\nu}_\text{df}(P_f,p';P_i;q_f) &= f(Q_i^2)(i\Hc^\mu_\mathrm{on}(P_f,p';P_s) - i\overline{\Hc}^\mu(P_u,p';P_i))\,.\label{eq:WT.qiTsimp}
\end{align}
These expressions can be exploited to obtain an expression of $\Tc_{\df}^{\mu\nu}$ in terms of simpler amplitudes whenever one of the external photon momenta vanishes.
Let us now specialize to the case of vanishing final photon momentum, expand the right hand side of Eq.~\eqref{eq:qfTsimp} around $q_{f,\mu}=0$, and keep only the first order term. This can be equated to obtain the $\Tc_{\df}^{\mu\nu}(P_f,p';P_i;0)$ amplitude. For that we begin by showing the dependence on $q_{f,\mu}$ explicitly
\begin{multline}
q_{f,\mu} i\Tc^{\mu\nu}_{21,\mathrm{df}}(P_f,p';P_i;q_f)
= f(Q_f^2)\sum_{\ell m} \sqrt{4\pi}Y_{\ell m}(\hat{p}^{\prime\star}_f)\Big(
i\mathcal{M}_\ell((P_f+q_f)^2)\mathcal{A}^\nu_{21,\ell m}(P_f+q_f,P_i)
\qty(\frac{p^{\prime\star}_{s}}{q^\star_{s}})^\ell
\\ 
- 
i\mathcal{M}_\ell(s_f)\mathcal{A}^\nu_{21,\ell m}(P_f,P_i-q_f)
\Big)\,,
\end{multline}
where $q^\star_{s}$ is the two-particle relative momentum in the CM frame of the intermediate s-channel.
The momentum magnitudes $p^{\prime \star}_{s}$ and $q^\star_{s}$ depend implicitly on $q_f$ since $P_s=P_f+q_f$. The first order expansion of the amplitudes $\Mc$ and $\Ac^\nu$ is straightforward, and the expansion of the barrier factors can be shown to be equal to
\begin{align}
        \qty(\frac{p^{\prime\star}_{s}}{q^\star_{s}})^\ell
&\equiv 
\left(
\frac{ 
\sqrt{
\frac{(p'\cdot (P_f + q_f))^2}{(P_f + q_f)^2}-m^2
}}
{\sqrt{\frac{(P_f+q_f)^2}{4} -m^2}}
\right) ^\ell
\\
& = 1- \frac{\ell}{ q_f^{\star 2}}q_{f,\mu} ( P_f-p')^\mu
+\mathcal{O}(q_{f,\mu}q_{f,\nu})\,,
\end{align}
where $q^{\star}_f$ is the relative momentum of the final two-particle state. 
In the limit that external photon momentum $q_{f,\mu}=0$ we find that
\begin{multline}
\Tc^{\mu\nu}_\text{21,df}(P_f,p';P_i;0)
=f(0)\sum_{\ell m}\sqrt{4\pi}Y_{\ell m}(\hat{p}_f^{\prime \star})\Big(
2P_f^\mu \pdv{\Mc_\ell(s_f)}{s_f}\Ac^\nu_{21, \ell m}(P_f,P_i) 
\\
+ 
\Mc_\ell(s_f)\qty(\pdv{P_f^\mu}+\pdv{P_i^\mu} )\Ac^\nu_{21, \ell m}(P_f,P_i) 
-
\frac{\ell}{ q_f^{\star 2}}  ( P_f -p')^\mu
\Mc_\ell(s_f)\Ac^\nu_{21,\ell m}(P_f,P_i)
\Big)\,.
\end{multline}
By means of crossing symmetry, or by repeating the previous steps beginning with Eq.~\eqref{eq:WT.qiTsimp} instead of Eq.~\eqref{eq:qfTsimp} we find, in limit that $q_{i}=P_f +q_f - P_i$ vanishes, the divergent free amplitude is equal to
\begin{multline}
\label{eq:WT.Tdfatq0}
\Tc^{\mu\nu}_{21,\df}(P_f,p';P_i;P_i-P_f)
=
f(0)\sum_{\ell m}\sqrt{4\pi}Y_{\ell m}(\hat{p}_f^{\prime \star})
\Big(
2P_f^\nu \pdv{\Mc_\ell(s_f)}{s_f}\Ac^\mu_{21, \ell m}(P_f,P_i) \\ + 
\Mc_\ell(s_f)
\qty(\pdv{P_f^\nu}+\pdv{P_i^\nu} )\Ac^\mu_{21, \ell m}(P_f,P_i) 
- 
\frac{\ell}{ q_f^{\star 2}}  ( P_f -p')^\nu
\mathcal{M}_\ell(s_f)\Ac^\mu_{21,\ell m}(P_f,P_i)
\Big)
\,.
\end{multline}
These equations show that the Ward identity constraints the short-distance piece $\Bc_{21}$ to be given in terms of simpler amplitudes whenever one of the photon momenta vanishes.

%%%%%%%%%%%%%%%%%%%%%%%%%%%%%%%%%%%%%%%%%%%
%  Section: Derivation of On-Shell Representations
%%%%%%%%%%%%%%%%%%%%%%%%%%%%%%%%%%%%%%%%%%%
\section{Derivation of On-Shell Representations}
\label{sec:Der_Onshell}

Here we present the derivation for the two Compton-like amplitudes presented in Sec.~\ref{sec:main}. 
We review first the known on-shell relation for the Compton amplitude of a single hadron, written in Eq.~\eqref{eq:T11_on-shell}.
Then, we present our new result for the Compton-like amplitude involving two hadrons, the $1+\Jc\to 2 + \Jc$ process, given in Eq.~\eqref{eq:T21_on-shell}.
In both cases, the final hadronic state has an outgoing total momentum $P_f$, and the final current is extracting momentum $q_f$, while the initial hadronic state has an incoming total momentum $P_i$, and the initial current injects momentum $q_i = P_f + q_f - P_i$.

In order to simplify the derivation, we will make a set of assumptions which will be lifted in Sec.~\ref{sec:general_case}. First, we assume that only one channel composed of two particles may be kinematically open. We will assume that the particles carry the same mass, which we will label as $m$, but only one of these can couple to the external current. Although in this sense, the particles are distinguishable, we will introduce a symmetry factor, $\xi$, which is defined to be $1/2$ if the particles are identical and $1$ otherwise. This will serve as book-keeping for when we lift these assumptions. 

In what follows, we assume that the external particles can couple to the external currents. As a result, for the simple Compton amplitude $1+\Jc\to 1+\Jc$, we expect a simple single-particle pole contributing to the amplitude. 

Because the fully-dressed single particle propagator only depends on the masses of the particles, we will label it simply as $\Delta(k)$ for all particles of momentum $k$. The simple pole contribution of the propagator is labeled as $D(k^2)$ and defined in Eq.~\eqref{eq:D_prop}. 
The non-analytic pieces of loop integrals emerge from these simple poles. Near the single-particle pole, the difference between $\Delta$ and $D$ is a smooth function, whose contribution will be absorbed into smooth kernels in the derivation below.
In order to simplify the notation further, we will introduce a symbol for the product of two propagators, $\Delta^{(2)}$, defined by 
\begin{align}
    \,\Delta^{(2)}(P,k) \equiv i\Delta(k)\, i
\Delta(P-k).
\end{align}

The Compton-like amplitudes have contributions from direct and exchange diagrams, which in the case of identical currents are related via crossing symmetry. Throughout the derivations that follow, we will consider the direct contribution, where the intermediate states have momentum $P_s$, i.e.\ the sum of the initial state momentum and that of the incoming current $B$. This $s$-channel contribution to the amplitude will be made explicit by introducing $AB$ subscripts in the various building blocks. It is relatively straightforward to obtain the $u$-channel contributions by replacing $A\leftrightarrow{}\!B$ and changing the corresponding kinematic dependence of the amplitudes, i.e.\ $q_i\leftrightarrow{}\!-q_f$ and $s\leftrightarrow\!u$. This will be done at the end of the derivation.

As is evident from the final expressions presented in Sec.~\ref{sec:main}, the singularity structure of these amplitudes depends on the amplitudes associated with physical sub-processes, $\mathcal{M}$, $w_{\mathrm{on}}$, $\mathcal{H}$, and $\mathcal{W}$. In Sec.~\ref{sec:main} we provided the on-shell representation of these. Here we provide the expressions of the integral equations for the off-shell $\Mc$ and $\Hc$ amplitudes, since these will be used in the subsequent derivation,~\footnote{The expression for $\Hc$ was first given in Ref.~\cite{Briceno:2014uqa}.} 
\begin{align}
\label{eq:M_dse}
i\Mc(p',{p})
&=
i\Kc_{0}(p',{p})
+
\xi\int \! \frac{\diff^{4}k}{(2\pi)^{4}} i\Mc(p',k)
\,\Delta^{(2)}(P_f,k)\,
i\Kc_{0}(k,p) \, ,
\end{align}
%%%%%%%%%%%%%%%%%%%%%%%%%%%%%%%%%%%%
\begin{align}
\label{eq:H_dse}
i\Hc^A(P_f,p';P_i) &= i\Hb^A_0(P_f,p';P_i) +
\xi\int
\frac{\diff^4k}{(2\pi)^4}
i\Mc(p',k)\,\Delta^{(2)}(P_f,k)\,i\Hb^A_0(P_f,k;P_i)\,,
\end{align}
where the momentum flowing through $\Mc$ and $\Kc_0$ is labeled as $P_f$ but it has been left implicit in its arguments. The external momenta $p$ and $p'$ denote the off-shell momenta of particle two in the initial and final state respectively. Beyond the dependence on these external momenta, the fact that the amplitudes are off-shell is left implicit in here. Once the amplitudes $\Mc$ and $\Hc$ have been partial wave projected, which is implicitly done with the ``on'' subscript or the barred operator $\overline{\Hc}$, they can be understood as being the on-shell amplitude.

In Eq.~\eqref{eq:H_dse} we separated the kinematic variables of $\Hc$ associated with the initial and final states by a semicolon. When the particle carrying the momentum $p'$ goes on-shell and the subsequent amplitude is partial wave-projected, the dependence on $p'$ will be trivial and is omitted from the subsequent expressions.

The kernels $\Kc_0$ and $\Hb_0$ are smooth functions up to the first unaccounted physical threshold. For now, this inelastic threshold could include a second two-particle channel, but after our generalization this must include three or more particles. This will be true for all the kernels considered in the following derivation, which will be labeled by boldfaced capital letters. 

To further simplify the following derivation we will introduce a compact notation for the functions and integrals considered. When we first write down the integral equation considered, we will show all kinematic and integration variables explicitly. After having written these expressions, we will proceed to manipulate each element by leaving its kinematic arguments and integration measure suppressed. The measure will be denoted using 
\begin{align}
    \int \frac{\diff^4k}{(2\pi)^4}
\longrightarrow \int_k.
\end{align}
Using this notation, the integral equation for $\Hc$, Eq.~\eqref{eq:H_dse}, can be rewritten as,
\begin{align}
\label{eq:H_dse_v2}
i\Hc^A  &= i\Hb^A_0 +
\xi\int_k
i\Mc \cdot \Delta^{(2)} \cdot i\Hb_0^A \,.
\end{align}
The dots separating each element remind the reader that these are functions of the internal flowing momenta. 

Finally, the off-shell extension of the one-particle matrix elements will be labeled by, 
\begin{equation}
    w^{A}(k_{f},k_{i})
=
\sum_{j} K_{j}^{A}(k_{f},k_{i})f_{j}(Q^{2},k_{f}^2,k_{i}^2),
\end{equation}
where $K_{j}$ are kinematic prefactors and $f_{j}(Q^{2},k_{f}^2,k_{i}^2)$ are the generalized off-shell form factors. As first described in Ref.~\cite{Baroni:2018iau}, one can recover the standard on-shell form factors, $ f_j(Q^2) $, by fixing the external momenta on-shell, i.e. $k_f^2=k_i^2=m^2$. We refer the reader to App.~\ref{sec:App.osproj} for a discussion about the prescription for the on-shell expansion of the form factors.

%%%%%%%%%%%%%%%%%%%%%%%%%%%%%%%%%%%%%%%%%%%
%  Subsection: T11 amplitude
%%%%%%%%%%%%%%%%%%%%%%%%%%%%%%%%%%%%%%%%%%%
\subsection{The \texorpdfstring{$1+\Jc\to 1+\Jc$}{1Jto1J} Compton amplitude}
\label{sec:1Jto1J}

\begin{figure}[t]
\begin{center}
\includegraphics[width=.7\textwidth]{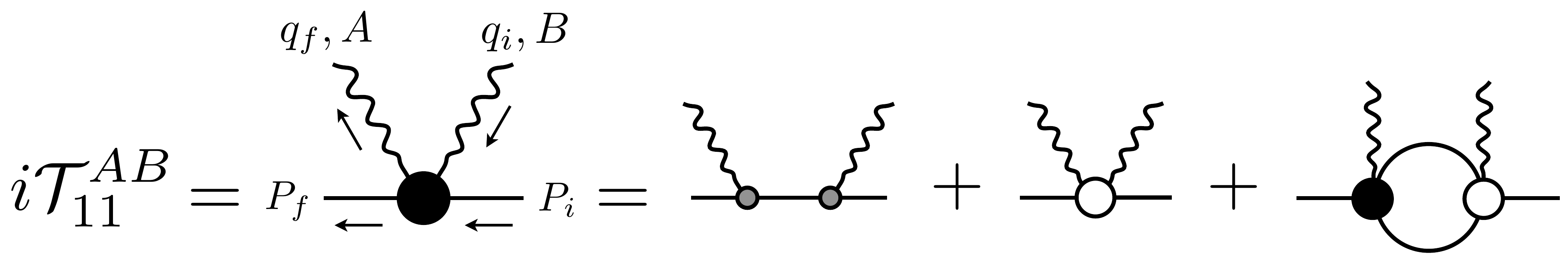}
\caption{ 
Self-consistent integral equation for the Compton amplitude $i\Tc^{AB}_{11}$, only the direct channel diagrams are shown. The grey circles represent the one-body matrix element, the open white circles represent the kernels $i\Tb^{AB}_{11,0}$ and $i\Hb_0^A$ respectively, which contain all the two-particle irreducible diagrams in the $P_s^2=s$ channel. 
}
\label{fig:T11}
\end{center}
\end{figure}

We begin by reproducing the derivation of the on-shell projection for $\Tc_{11}$ first presented in Ref.~\cite{Briceno:2019opb}. The steps closely resemble those presented in Ref.~\cite{Briceno:2020vgp} for amplitudes involving a single current insertion.

Summing to all orders in the strong interaction, depicted in Fig.~\ref{fig:T11}, we find that the $s$-channel contributions to $\Tc_{11}$ can be written as
\begin{multline}
    \label{eq:1Jto1J.T11_dse}
    i\Tc_{11}^{AB}(P_f,P_i,q_f) 
    =
    iw^{A}(P_f,P_s) i\Delta(P_s) iw^{B}(P_s,P_i) \\%\nn \\[5pt]
    + 
    i\Tb_{11,0}^{AB}(P_f,P_i,q_{f}) 
    + \xi \int\!\frac{\diff^4 k}{(2\pi)^4} \,i\Hb_{0}^{A}(P_f;P_s,k) 
    \Delta^{(2)}(P_s,k)
    i\Hc^{B}(k,P_s;P_i),
\end{multline}
where we remind the reader $P_{s}=P_{f}+q_{f}=P_{i}+q_{i}$ such that $P_{s}^{2}=s$. The superscripts $AB$ are to remind the reader that this is the contribution due to direct diagrams only. 

The kernel $\Tb_{11,0}$ couples one-particle states via two-current insertions. By making the single-particle poles and two-particle cuts explicit, $\Tb_{11,0}$ is defined to be one- and two-particle $s$-channel irreducible, and consequently it is a smooth, non-singular function in the kinematic region of interest.
 
The pole term can be put into an on-shell form by expanding $w$ about the on-shell point for the internal propagator.
The remaining short-distance contributions can be absorbed into a single function, which we will denote as $\Tb_{11,\alpha}$,
\begin{align}
\label{eq:1Jto1.wDw_proj}
    iw^{A}\cdot i\Delta\cdot iw^{B}  
    =
    i\Tb_{11,\alpha}^{AB}(P_f,P_i,q_f) %\nn\\[5pt]
    +
    iw^{A}_{\mathrm{on}}(P_f,P_s)\, iD(s)\, iw^{B}_{\mathrm{on}}(P_s,P_i).
\end{align}
A supplementary discussion about the on-shell expansion process of kernel $w$ can be found in App.~\ref{sec:App.osproj}.

For the final term in Eq.~\eqref{eq:1Jto1J.T11_dse} we use a similar procedure as that shown in Ref.~\cite{Briceno:2020vgp}, where we start by substituting the integral relation for $\Hc$, Eq.~\eqref{eq:H_dse}, such that we are left with integrals representing loops with smooth, non-singular kernels on each vertex. As reviewed in some detail in Appendix~\ref{sec:App.Iden}, we can then separate out the singular pieces of these integrals by taking advantage of the fact that in our limited kinematic region the only singularities that may occur come from the intermediate two-particles state going on-shell. Making this separation and partial-wave projecting the kernels to complete the integration over the singular piece, we find
\begin{align}
    \label{eq:1Jto1.loop_id}
    \xi \int_k  
    \, i\Hb_{0}^{A}\cdot  
    \Delta^{(2)}\cdot
    i\Hc^{B} 
    & =
    \sum_{\ell,m_{\ell}} i\Hb_{0;\ell m_{\ell}}^{A}(P_f,P_s) \rho \, i\Hc_{\ell m_{\ell}}^{B}(P_s,P_i) 
    + i \Tb_{11,1}^{AB}(P_f,P_i,q_f) \nn \\
    & + \xi \int \! \frac{\diff^4 k}{(2\pi)^{4}} \, i\Hb_{1}^{A}(P_f;P_s,k) \, \Delta^{(2)}(P_s,k) i\Hc^{B}(P_s,k;P_i) \, .
\end{align}
This on-shell separation results in two new kernels $\Hb_1$ and $\Tb_{11,1}$, both of which come from the off-shell contributions of the loop integral. $\Hb_1$ comes with a kernel $\Kc_0$ in one vertex and $\Hb_0$ in the other, while $\Tb_{11,1}$ has kernels $\Hb_0$ on each vertex. The last term of Eq.~\eqref{eq:1Jto1.loop_id} has the same structure as the left hand side, therefore we can repeat the step shown in Eq.~\eqref{eq:1Jto1.loop_id}, and generate an infinite number of iterations for $\Tb_{11,j}$ and $\Hb_{j}$ for $j \in \mathbb{N}$.
After summing all terms, we arrive at the on-shell expression
\begin{align}
\label{eq:T11_T11b}
    i\Tc_{11}^{AB}(P_f,P_i,q_f) 
    & =
    iw_{\mathrm{on}}^A(P_f,P_s) \,iD(s) \,iw_{\mathrm{on}}^{B}(P_s,P_i) \nn\\[5pt]
    & +
    i\Tb_{11}^{AB}(P_f,P_i,q_f) + \sum_{\ell,m_{\ell}}
    i\Hb_{\ell m_{\ell}}^{A}(P_f,P_s)
    \, \rho \,
    \Mc_{\ell}(s) \, i\Ac_{21,\ell m_{\ell}}^{B}(P_s,P_i) \, ,
\end{align}
where $\Tb_{11}$ includes the sum of all iterated $\Tb_{11,j}$ kernels as well as $\Tb_{11,\alpha}$, $\Hb$ is the sum of all $\Hb_j$ kernels, and we use Eq.~\eqref{eq:H_on-shell} to rewrite $\Hc$ in terms of $\Mc$ and $\Ac_{21}$.

As discussed in Ref.~\cite{Briceno:2020vgp}, if the $K$-matrix has unphysical poles, $\Hb$ will contain these same poles. This can be made explicit by writing it as 
\begin{align}
\label{eq:Hb_def}
\Hb_{\ell m_{\ell}}^{A}(P_f,P_s)
=
\Ac_{12,\ell m_{\ell}}^{A}(P_f,P_s)
\Kc_{\ell}(s).
\end{align}
These poles arise by the all-orders summation of the smooth contribution to the $s$-channel loop integrals. These must be absent in $\Tc_{11}$ in order to assure that it remains analytic except for singularities required by unitarity. Any unphysical pole present in the last term of Eq.~\eqref{eq:T11_T11b} have to be canceled exactly by unphysical $K$-matrix poles present in $\Tb_{11}$. We make these explicit by rewritting $\Tb_{11}$ as 
\begin{align}
    \Tb_{11}^{AB}(P_f,P_i,q_f) 
    = 
    \Bc_{11}^{AB}(P_f,P_i,q_f)  + \sum_{\ell,m_{\ell}} \Ac_{12,\ell m_{\ell}}^{A}(P_f,P_s) \Kc_{\ell}(s) \Ac_{21,\ell m_{\ell}}^{B}(P_s,P_i) \, ,
\end{align}
where $\Bc_{11}$ is a real and smooth function in the restricted kinematic domain.
Inserting this as well as Eq.~\eqref{eq:Hb_def} for $\Hb$ into Eq.~\eqref{eq:T11_T11b}, we can write our final expression for the $s$-channel contributions
\begin{align}
\label{eq:1Jto1J.OnShell}
    i\Tc_{11}^{AB}(P_f,P_i,q_f) 
    & = 
    iw_{\mathrm{on}}^{A}(P_f,P_s) iD(s) iw_{\mathrm{on}}^{B}(P_s,P_i) \nn \\[5pt]
    & + i \Bc_{11}^{AB}(P_f,P_i,q_f) + \sum_{\ell,m_{\ell}} \Ac_{12,\ell m_{\ell}}^{A}(P_f,P_s) i\Mc_{\ell}(s) \Ac_{21,\ell m_{\ell}}^{B}(P_s,P_i),
\end{align}
where on-shell expression of $\Mc$ in Eq.~\eqref{eq:M_on-shell} was used to further simplify the last term.
Below the two-particle threshold, the last term on the right hand side of Eq.~\eqref{eq:1Jto1J.OnShell} becomes a smooth analytic contribution, up to possible bound state poles in the two-particle channel, which are encoded in $\Mc$.

As promised, we can now easily include the contribution from the exchange diagrams from Eq.~\eqref{eq:1Jto1J.OnShell} by swapping the $A$ and $B$ indices, and changing $q_f\to -q_i$, which results in $s\to u$. Adding these two contributions, and again using the previously introduced notation $P_{u}=P_{f}-q_{i}=P_{i}-q_{f}$ such that $P_{u}^{2}=u$, we arrive at our final expression for $\Tc_{11}$,
\begin{align}
    i\Tc_{11}(P_{f},P_{i},q_{f})
    &=
    iw_{\mathrm{on}}^{A}(P_f,P_s) iD(s) iw_{\mathrm{on}}^{B}(P_s,P_i)
    +
    iw_{\mathrm{on}}^{B}(P_f,P_u) iD(u) iw_{\mathrm{on}}^{A}(P_u,P_i) \nn \\[5pt]
    & + i \Bc_{11}^{AB}(P_f,P_i,q_f)+ i \Bc_{11}^{BA}(P_f,P_i,-q_i) \nn \\[5pt]
    & +
    \sum_{\ell,m_{\ell}}
    \left[ 
    \Ac_{12,\ell m_{\ell}}^{A}(P_f,P_s) i\Mc_{\ell}(s) \Ac_{21,\ell m_{\ell}}^{B}(P_s,P_i)
    + \Ac_{12,\ell m_{\ell}}^{B}(P_f,P_u) i\Mc_{\ell}(u) \Ac_{21,\ell m_{\ell}}^{A}(P_u,P_i)
    \right].
\end{align}
It is worth noting that this result agrees with the expression given in Eq.~(53) of Ref.~\cite{Briceno:2019opb} for the Compton scattering amplitude.

%%%%%%%%%%%%%%%%%%%%%%%%%%%%%%%%%%%%%%%%%%%
%  Subsection: T21 amplitude
%%%%%%%%%%%%%%%%%%%%%%%%%%%%%%%%%%%%%%%%%%%
\subsection{The \texorpdfstring{$1 + \Jc\to 2 + \Jc$}{1Jto2J} Compton amplitude}
\label{sec:1Jto2J}

Having derived the on-shell representation for the case of the standard Compton amplitude we now move on to the derivation of our main result where we may have two hadrons in either the initial or final state but not both. We  denote this class of Compton-like amplitudes as $\Tc_{21}$ where the subscript tells us the number of hadrons in the final/initial state respectively. 

To begin, we again consider the case of the direct contributions, i.e.\ those appearing as $s$-channel intermediate states. From here we can then split the amplitude into two sets, one which contains kernels involving 1-body interactions, and the other which depends on a new short-distance kernel ($\Tb_{0|0}$) which, as with all other kernels, is smooth and non-singular in the kinematic region of interest. Labelling these as $i\Tc_{1\rm{B}}$ and $i\Tc_{\cancel{1\rm{B}}}$, respectively, we have
\begin{align}
i\Tc_{21}^{AB}
&=
i\Tc_{21,\cancel{1\rm{B}}}^{AB} + i\Tc_{21,1B}^{AB}.
\label{eq:T21_twoterms}
\end{align}
In Secs.~\ref{sec:T21no1B} and \ref{sec:T211B}, we give the governing equations of these terms respectively. 

\begin{figure}[t]
\begin{center}
\includegraphics[width=.7\textwidth]{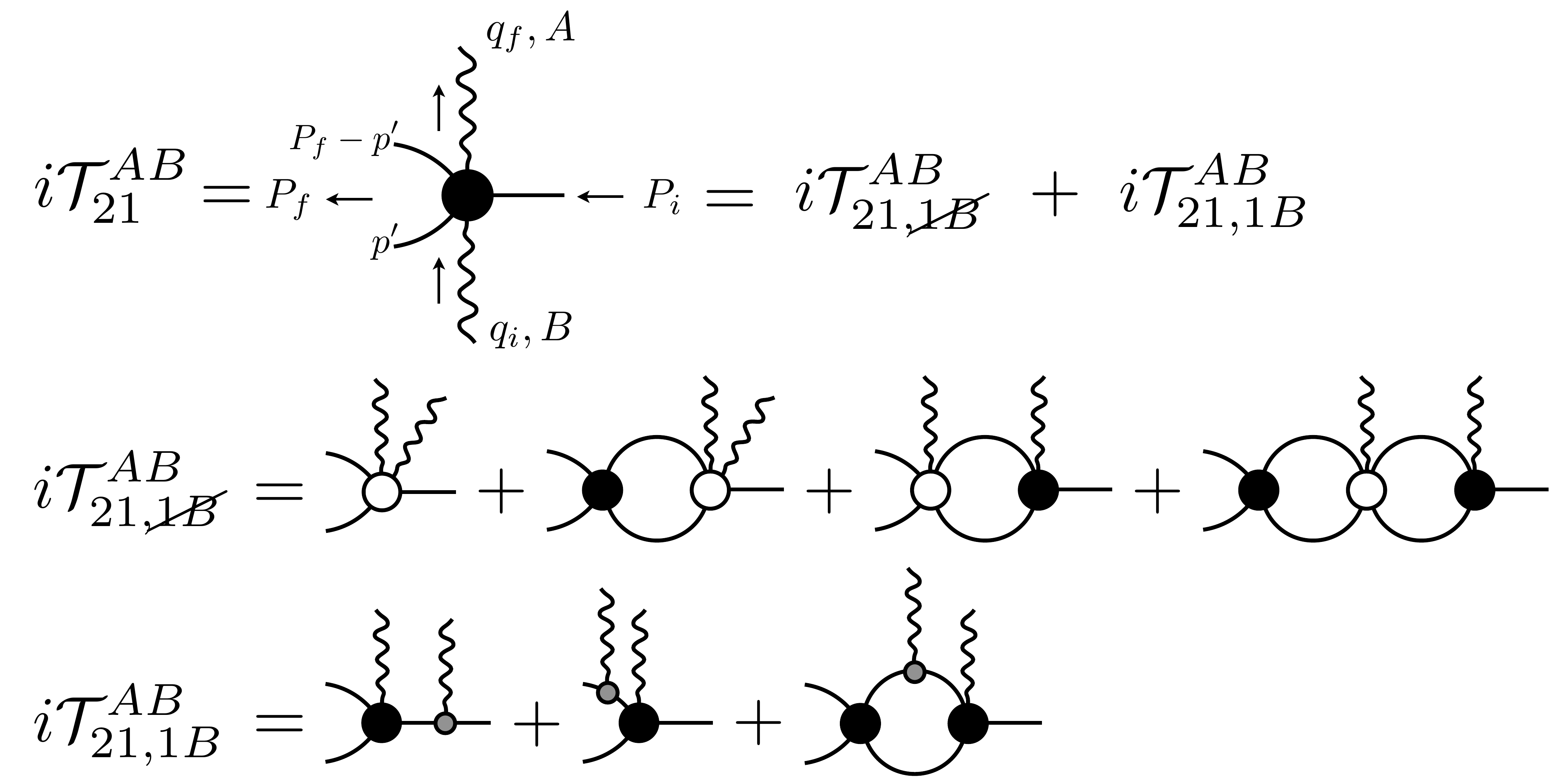}
\caption{
Diagrammatic representation of the transition amplitude $i\Tc_{21}^{AB}$. Most of the building blocks were previously defined in Figs.~\ref{fig:BuildingBlocks} and \ref{fig:T11}, with the exception of the white open circle connecting $1$ and $2$ hadronic states via the insertion of two local current, which is the diagrammatic representation of $\Tb_{0|0}^{AB}$.
}
\label{fig:T21}
\end{center}
\end{figure}

%%%%%%%%%%%%%%%%%%%%%%%%%%%%%%%%%%%%%%%%%%%
%  Subsubsection: No 1-body amplitude
%%%%%%%%%%%%%%%%%%%%%%%%%%%%%%%%%%%%%%%%%%%
\subsubsection{No 1-body amplitude}
\label{sec:T21no1B}

We begin by first analyzing the $\Tc_{21,\cancel{1\rm{B}}}$ amplitude, shown diagrammatically in Fig.~\ref{fig:T21}. To all orders in the strong interaction, this can be written as,
\begin{multline}
\label{eq:Cno1B_dse}
i\Tc_{21,\cancel{1\rm{B}}}^{AB}(P_{f},p';P_{i};q_{f})
=
i\Tb_{0|0}^{AB}(P_{f},p';P_{i};q_{f})
+
\xi \int \frac{\diff^{4}k}{(2\pi)^{4}}
i\Mc(p',k)  
\,\Delta^{(2)}(P_f,k)\,
 i\Tb_{0|0}^{AB}(P_{f},k;P_{i};q_{f})
\\+
\xi \int \frac{\diff^{4}k}{(2\pi)^{4}}
i\Wb^A_{0|0}(P_{f},p';P_{s},k) 
\, \Delta^{(2)}(P_s,k)\,
 i\Hc^B(P_{s},k;P_{i})
\\+
\xi \int \frac{\diff^{4}k'}{(2\pi)^{4}}
\xi \int \frac{\diff^{4}k}{(2\pi)^{4}}
i\Mc(p',k') 
\,\Delta^{(2)}(P_f,k')\,
i\Wb_{0|0}^{A}(P_{f},k';P_{s},k) 
\,\Delta^{(2)}(P_s,k)\,
i\Hc^{B}(P_{s},k;P_{i}),
\end{multline}
where we have introduced two new short distance kernels, $\Wb_{0|0}$ and $\Tb_{0|0}$, involving one and two current insertions, respectively. The former was introduced in Ref.~\cite{Briceno:2020vgp} for deriving the expression for $\Wc_\df$, given in Eq.~\eqref{eq:Wdf_def}.  

In Eq.~\eqref{eq:Cno1B_dse} we separated the kinematic variables of $\Tc_{21,\cancel{1\rm{B}}}$ and $\Wb_{0|0}$ associated with the final state, the initial state, and one of the currents with semicolons. As was the case with $\Hc$, we will only keep the semicolons for amplitudes that are off-shell and/or have not been partial-wave projected. This notation will be used throughout. 

The classes of integrals are identical to the ones considered in Ref.~\cite{Briceno:2020vgp} in the context of the $\Wc$ amplitude. As previously mentioned, in Appendix~\ref{sec:App.Iden} we provide the key identities needed to isolate the singularities of these. Using Eq.~\eqref{eq:App.FSIfin}, we can rewrite the first two terms of Eq.~\eqref{eq:Cno1B_dse} to isolate the phase space singularities
 \begin{align}
 \label{eq:Cno1B_eq1}
i\Tb_{0|0}^{AB}
+
\xi\int_k  i\Mc\cdot  
\,\Delta^{(2)}
\cdot i\Tb_{0|0}^{AB}
=
i\Tb_{\infty|0}^{AB}(P_f,p';P_i;q_f) +
\sum_{\ell, m_\ell}i\Mc_\ell(p') \, \rho \, i\Tb_{\infty|0,\ell m_\ell}^{AB}(P_f,P_i,q_f),
\end{align}
where $\Tb_{\infty|0}$ includes the sum over an infinite number of smooth iterated kernels $\Tb_{j|0}$.

For the third term in Eq.~\eqref{eq:Cno1B_dse} we use the integral equation for $\Hc$, Eq.~\eqref{eq:H_dse}, and Eq.~\eqref{eq:App.MSIsol}, to rewrite it as
\begin{equation}
 \label{eq:Cno1B_eq2}
\xi \int_k
i\Wb^A_{0|0}\cdot \Delta^{(2)} \cdot i\Hc^B 
=
\sum_{j=1}^\infty i\Tb^{AB}_{0|j}(P_f,p';P_i;q_f)+
\sum_{\ell,m_\ell}i\Wb^A_{0|\infty,\ell m_\ell}(P_{f},p';P_s) \, \rho \, i\Hc^B_{\ell m_\ell}(P_s,P_{i})\,.
\end{equation}
Finally, the last term in Eq.~\eqref{eq:Cno1B_dse} can be added to Eq.~\eqref{eq:Cno1B_eq2} using Eq.~\eqref{eq:App.FSIfin},
\begin{multline}\label{eq:Cno1B_eq3}
    \xi \int_k
i\Wb^A_{0|0}\cdot \Delta^{(2)} \cdot i\Hc^B 
+ \xi \int_{k'}
\xi \int_k
i\Mc\cdot
\,\Delta^{(2)}\cdot
i\Wb_{0|0}^{A}\cdot
\,\Delta^{(2)}\cdot
i\Hc^{B}
=
\sum_{j=1}^\infty i\Tb^{AB}_{\infty|j}(P_f,p';P_i;q_f)\\+
\sum_{\ell,m_\ell}i\Wb^A_{\cancel{1\rm{B}},\ell m_\ell}(P_{f},p';P_s) \, \rho \, i\Hc^B_{\ell m_\ell}(P_s,P_{i})
\\+
\sum_{\ell', m_\ell'}i\Mc_{\ell'}(p') \,\rho
\left(
\sum_{j=1}^\infty
i\Tb^{AB}_{\infty|j,\ell' m_\ell'}(P_f,P_i,q_f)+
\sum_{\ell,m_\ell}i\Wb^A_{\cancel{1\rm{B}};\ell'm_\ell';\ell m_\ell}(P_{f},P_s) \, \rho \, i\Hc^B_{\ell m_\ell}(P_s,P_{i})
\right),
\end{multline}
where we have introduced the function $\Wb^A_{\cancel{1\rm{B}}}$, defined as 
\begin{align}
    \Wb^A_{\cancel{1\rm{B}};\ell'm_\ell';\ell m_\ell} \equiv 
    \sum_{j',j=0}^\infty
i\Wb^A_{j'|j,\ell'm_\ell',\ell m_\ell}.
\end{align}

Adding Eqs.~\eqref{eq:Cno1B_eq1} and \eqref{eq:Cno1B_eq3} we arrive at the final expression for $i\Tc_{21,\cancel{1\rm{B}}}$, 
\begin{multline}
\label{eq:Cno1B_dse_v2}
i\Tc_{21,\cancel{1\rm{B}}}^{AB}
=
i\Tb_{\cancel{1\rm{B}}}^{AB}(P_f,p';P_i;q_f) +
\sum_{\ell, m_\ell}i\Mc_\ell(p') \,\rho\, i\Tb_{\cancel{1\rm{B}},\ell m}^{AB}(P_f,P_i,q_f)
\\
+  \sum_{\ell,m_\ell}i\Wb^A_{\cancel{1\rm{B}};\ell m_\ell}(P_{f},p';P_s) \, \rho \, i\Hc^B_{\ell m_\ell}(P_s,P_{i})
\\
+
\sum_{\ell', m_\ell'}i\Mc_{\ell'}(p') \,\rho\,
\sum_{\ell,m_\ell}i\Wb^A_{\cancel{1\rm{B}};\ell'm_\ell';\ell m_\ell}(P_{f};P_s) \, \rho \, i\Hc^B_{\ell m_\ell}(P_s,P_{i})\,,
\end{multline}
where we have added the smooth contributions into a single kernel
\begin{equation}
\Tb_{\cancel{1\rm{B}}}^{AB}(P_f,p';P_i;q_f)
=
\sum_{j',j=0}^\infty  \Tb^{AB}_{j'|j}(P_f,p';P_i;q_f)\,.
\end{equation}
Finally, we partial-wave project the final state so that the result simplifies to
\begin{equation}
\label{eq:Cno1B_dse_vf}
i\Tc_{21,\cancel{1\rm{B}},\ell m_\ell}^{AB}(P_{f},P_{i},q_{f})
=
(1+i\Mc_\ell(s_f) \,\rho)\Big(i\Tb_{\cancel{1\rm{B}},\ell m}^{AB}(P_f,P_i,q_f)
+  \sum_{\ell',m_\ell'}i\Wb^A_{\cancel{1\rm{B}};\ell m_\ell;\ell' m_\ell'}(P_{f},P_s) \, \rho \, i\Hc^B_{\ell' m_\ell'}(P_s,P_{i})
\Big)\,.
\end{equation}
This result is easy to understand. In the absence of the one-body couplings to the current, the amplitude does not have triangle singularities. As a result, the only source of singularities are due to $s$-channel bubble diagrams, which result in the $\rho$ cuts. 

It is worth noting that both $\Tb_{\cancel{1\rm{B}}}$ and $\Wb^A_{\cancel{1\rm{B}}}$ can have unphysical $K$-matrix poles. We make these explicit in our final expression for $\Tc_{21}$. Finally, Eq.~\eqref{eq:Cno1B_dse_vf} only includes contributions from the $s$-channel. As with $\Tc_{11}$, it is straightforward to include the contribution from the $u$-channel diagrams, which we will do once we have derived the on-shell representation for $\Tc_{21,1B}$.

%%%%%%%%%%%%%%%%%%%%%%%%%%%%%%%%%%%%%%%%%%%
%  Subsubsection: 1-body amplitude
%%%%%%%%%%%%%%%%%%%%%%%%%%%%%%%%%%%%%%%%%%%
\subsubsection{1-body amplitude}
\label{sec:T211B}

Having dealt with $\Tc_{21,\cancel{1\rm{B}}}$ we move on to the second term contributing to Eq.~\eqref{eq:T21_twoterms}, namely $\Tc_{21,1B}$. Its diagrammatic representation is shown in Fig.~\ref{fig:T21}, and its underlying equation can be written as, 
\begin{multline}
\label{eq:C1B_dse}
i\Tc_{21,1B}^{AB}(P_f,p';P_i;q_f)
=
i\Hc^{A}(P_f,p';P_s) i\Delta(P_s) iw^{B}(P_s,P_i)
+
iw^{A}(p'_f,p'_s) i\Delta(p'_s) i\Hc^{B}(P_s,p';P_i)
\\+
\int
\frac{\diff^4k}{(2\pi)^4}
i\Mc(p',k) \Delta^{(2)}(P_f,k)
iw^{A}(k_f,k_s) i\Delta(k_s)  i\Hc^{B}(P_s,k;P_i),
\end{multline}
where $p'_{f/s}\equiv P_{f/s}-p'$,  $k_{f/s}\equiv P_{f/s}-k$, and all other building blocks have been previously defined.

\begin{figure}[t]
\begin{center}
\includegraphics[width=.7\textwidth]{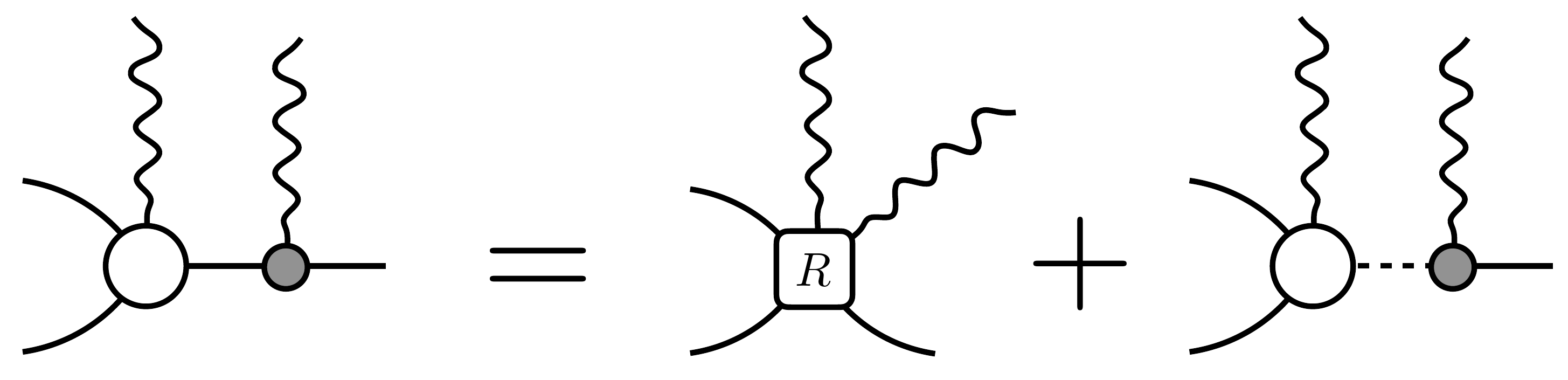}
\caption{ 
Diagrammatic form of the on-shell expansion shown in Eq.~\eqref{eq:T21_1B_sinpart}. The first term is a new smooth kernel which we define to be $\Tb_{0|R}^{AB}$ and the dotted line in the second term represents the pole piece of the propagator, $D$.
}
\label{fig:Right_OnShell}
\end{center}
\end{figure}

We begin by isolating the pole contribution of the propagator of the first term. To do this, we first use the definition of the transition amplitude, $\Hc^A$, given in Eq.~\eqref{eq:H_dse_v2} in terms of the $\Hb^A$ kernel. Next, we place the adjacent kernels on their mass shell. By replacing $\Hc^A$ with $\Hb^A_0$, we get the first contribution to $i\Hc^{A}\cdot i\Delta\cdot iw^{B}$,
\begin{align}\label{eq:T21_1B_sinpart}
i\Hb_0^{A}(P_{f},p';P_{s}) i\Delta(P_{s}) iw^{B}(P_{s},P_{i})
=
i \Tb_{0|R}^{AB}(P_{f},p';P_{i};q_{f})
+
i \mathbf{H}_{0,\mathrm{on}}^{A}(P_{f},p';P_{s}) iD({s}) iw^{B}_{\rm on}(P_{s},P_{i})\,,
\end{align}
where $\Tb_{0|R}$ is a new smooth function absorbing all off-shell effects, and the $R$ in the subscript is meant to remind us that the single-current coupling is taking place to the \emph{right} of the diagram. We illustrate this procedure diagramatically in Fig.~\ref{fig:Right_OnShell}. The subscript ``on'' in the kernel $\Hb_{\rm on}$ is to emphasize that even if $P_s$ is off shell, the energy-dependendent transition form factors within are to be projected on shell. Appendix~\ref{sec:App.osproj} provides further discussion about this procedure and an explicit treatment in the case of a conserved vector current.
Using the all orders definition of $\Hc$, Eq.~\eqref{eq:H_dse_v2}, we get,
\begin{align}
i\Hc^{A}\cdot i\Delta \cdot iw^{B}
&=
i\Hc^{A}_{ \mathrm{on}}(P_{f},p';P_{s}) iD({s}) iw^{B}_{ \mathrm{on}}(P_{s},P_{i})
+
i \Tb_{0|R}^{AB}(P_{f},p';P_{i};q_{f})\\
&\hspace{3.5cm}
+
\xi\int \frac{d^{4}k}{(2\pi)^{4}} 
i\Mc(p',k) \Delta^{(2)}(P_{f},k) i\Tb_{0|R}^{AB}(P_{f},k;P_{i};q_{f})\nn\\
%%%%%%%%%%%%%%%%%%%%%%%
&=
i\Hc^{A}_{ \mathrm{on}}(P_{f},p';P_{s}) iD({s}) iw^{B}_{ \mathrm{on}}(P_{s},P_{i})
+
i \Tb_{\infty|R}^{AB}(P_{f},p';P_{i};q_{f})\nn\\
&\hspace{3.5cm}+
\sum_{\ell, m_\ell} 
i\Mc_\ell(p') \, \rho \, i\Tb_{\infty|R,\ell m_\ell}^{AB}(P_{f},P_{i},q_{f}),
\label{eq:C1B_dse_term1_vf}
\end{align}
where we used Eq.~\eqref{eq:App.FSIfin} to write the second equality. 

For the second term in Eq.~\eqref{eq:C1B_dse}, we use the self-consistent integral equation for $\Hc$ to rewrite it as, 
\begin{multline}
iw^{A}\cdot i\Delta\cdot i\Hc^{B}
= 
iw^{A}(p'_f,p'_s) i\Delta(p'_s) i\mathbf{H}_0^{B}(P_{s},p';P_{i})
\\+
iw^{A}(p'_f,p'_s) i\Delta(p'_s)
\xi\int \frac{d^{4}k}{(2\pi)^{4}}
i\Mc(p',k) \Delta^{(2)}(P_{s},k) i\mathbf{H}_0^{B}(P_{s};P_{i},p).
\label{eq:C1B_dse_term2}
\end{multline}
 
\begin{figure}[t]
\begin{center}
\includegraphics[width=.7\textwidth]{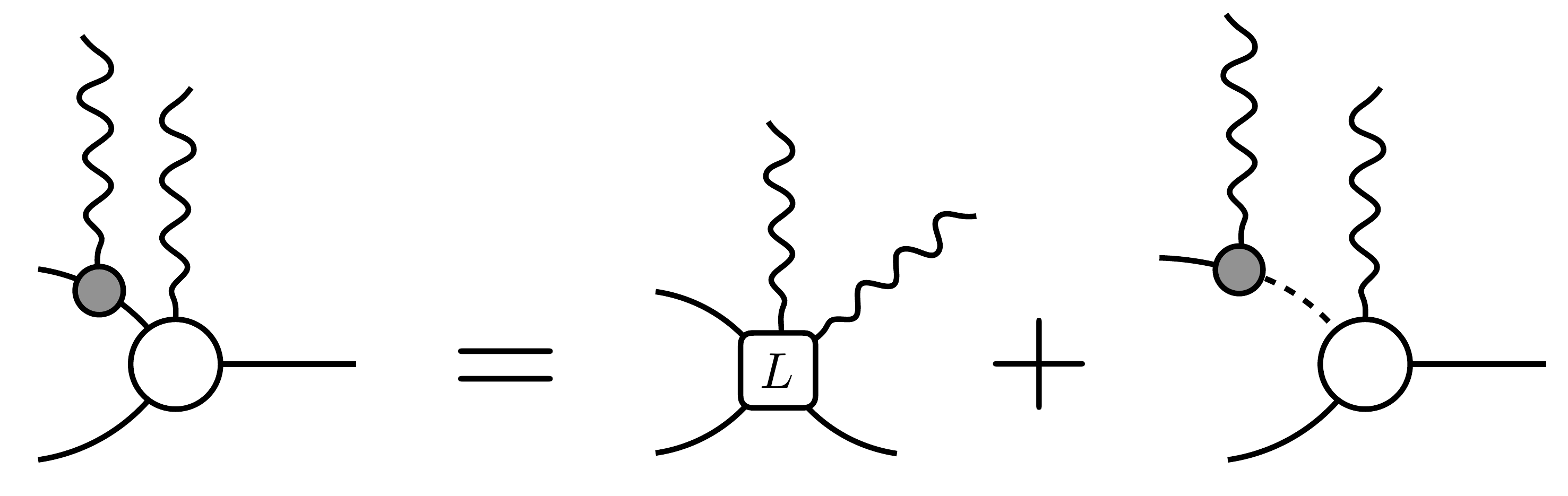}
\caption{ 
Diagrammatic form of the expansion shown in Eq.~\eqref{eq:T21.Hbar}. The first term is again a new smooth kernel we call $\Tb_{L|0}$ and the white open circle in the second term represents $\overline{\Hb}_{0}$ which is similar to $\Hb_{0}$ except that it contains the modified spherical harmonics $\Yc_{\ell m}$ as described in the text.
}
\label{fig:Left_OnShell}
\end{center}
\end{figure}

Once again, we can isolate the pole contribution by projecting the final state coupling to the $\Hb_0$ kernel on-shell. We do this by writing this kernel in terms of its CM coordinates and using spherical harmonics to parameterize the angular dependence. In order to assure that this procedure does not introduce spurious singularities, we use the modified spherical harmonics,  $\Yc_{\ell,m_{\ell}}$, defined in Eq.~\eqref{eq:sph_w_barrier}. With this, we find
\begin{align}\label{eq:T21.TL0def}
iw^{A}(p'_{f},p'_{s}) i\Delta(p'_{s}) i\mathbf{H}_0^{B}(P_{s},p';P_{i})
&= i\Tb_{L|0}^{AB}(P_f,p';P_i;q_f)
    +iw^{A}_{ \mathrm{on}}(p'_{f},p'_{s}) iD(p'^2_{s})
    \sum_{\ell, m_\ell}
    \Yc_{\ell m_\ell}({\mathbf p}_s^{\prime\star})
     i\Hb_{0,\ell m_\ell}^{B}(P_s,P_i)\,
    \\
    %%%%%%%%%%%%%%%%%%%%%%%%%%%
    &\equiv i\Tb_{L|0}^{AB}(P_f,p';P_i;q_f)
    +iw^{A}_{ \mathrm{on}}(p'_{f},p'_{s}) iD(p'^2_{s})i\overline{\Hb}_{0}^{B}(P_s,p';P_i)\,,\label{eq:T21.Hbar}
\end{align}
where in the last equality we have introduced a working definition for $\overline{\Hb}$. The smooth kernel $\mathbf{T}_{L|0}$ is similar to $\mathbf{T}_{0|R}$ except that the current is now coupling to the left of the diagram. This procedure is summarized in Fig.~\ref{fig:Left_OnShell}.

Next, we expand the second term of Eq.~\eqref{eq:C1B_dse_term2} by using the recursion relation~\eqref{eq:M_dse} such that only $\Kc_0$ kernels are next to the single particle intermediate state. We will use the on-shell expansion defined in Ref.~\cite{Briceno:2020vgp} 
\begin{equation}
    iw^{A}(p'_{f},p'_{s}) i\Delta(p'_{s}) i\Kc_0(p',k)
= i\Wb_{L|0}^{A}(P_f,p';P_s,k)
    +iw^{A}_{ \mathrm{on}}(p'_{f},p'_{s}) iD(p'^2_{s})i\overline{\Kc}_{0}(p',k)\,,\label{eq:T21.Kbar}
\end{equation}
where $\overline{\Kc}_0$ is defined in analogous way to $\overline{\Hb}_0$. Once all the barred kernels have been grouped together we recover the barred transition amplitude 
\begin{align}
    i\overline{\Hc}^B(P_s,p';P_i) 
    &\equiv i\overline{\Hb}_{0}^{B} + \xi\int_k i\overline{\Kc}_{0}\cdot\Delta^{(2)}\cdot i\Hb_{0}^{B}
    + \xi\int_k\xi\int_{k'} i\overline{\Kc}_{0}\cdot\Delta^{(2)}\cdot i\Mc \cdot\Delta^{(2)}\cdot i\Hb_{0}^{B}\\
    &=\sum_{\ell, m_\ell}
    \Yc_{\ell m_\ell}({\mathbf p}_s^{\prime\star})i\Hc^B_{\ell m_\ell}(P_s,P_i)\,, \label{eq:Hoverldef}
\end{align}
where the last equality is a consequence of the recursive definition of Eq.~\eqref{eq:H_dse} and the definition of a barred kernel.

This allows us to rewrite Eq.~\eqref{eq:C1B_dse_term2} in the following manner
\begin{multline}
iw^{A}\cdot i\Delta\cdot i\Hc^{B}
= iw^A_{ \mathrm{on}}(p'_f,p'_s) iD(p'^2_s) i\overline{\Hc}^B(P_s,p';P_i) 
\\
+ i\Tb_{L|0}^{AB}(P_f,p';P_i;q_f) + \xi\int \frac{\diff^4k}{(2\pi)^4} i \Wb^A_{L|0}(P_f,p';P_s,k) \Delta^{(2)}(P_{s},k) i\Hc^B(P_s,k;P_i)\,.
\end{multline}
To finish the simplification of Eq.~\eqref{eq:C1B_dse_term2} we apply Eq.~\eqref{eq:App.MSIsol} to isolate the singularities of the remaining integral to find
\begin{multline}
iw^{A}\cdot i\Delta\cdot i\Hc^{B}
= iw^A_{ \mathrm{on}}(p'_f,p'_s) iD(p'^2_s) i\overline{\Hc}^B(P_s,p';P_i) 
\\
+ i\Tb^{AB}_{L|\infty}(P_f,p';P_i;q_f) +  
\sum_{\ell, m_\ell}i\Wb^A_{L|\infty,\ell m_\ell}(P_f,p';P_s) \,\rho\, i\Hc^B_{\ell m_\ell}(P_s,P_i)\,,
\label{eq:C1B_dse_term2_vf}
\end{multline}
where we defined the kernels $\Tb_{L|j}$ kernels following Eq.~\eqref{eq:App.Ijdefin} by replacing $\Ab_j\to\Wb_{L|j}$, $\Bb_0\to\Hb_0$, and $\Ib_j\to \Tb_{L|j}$.

\begin{figure}[t]
\begin{center}
\includegraphics[width=.9\textwidth]{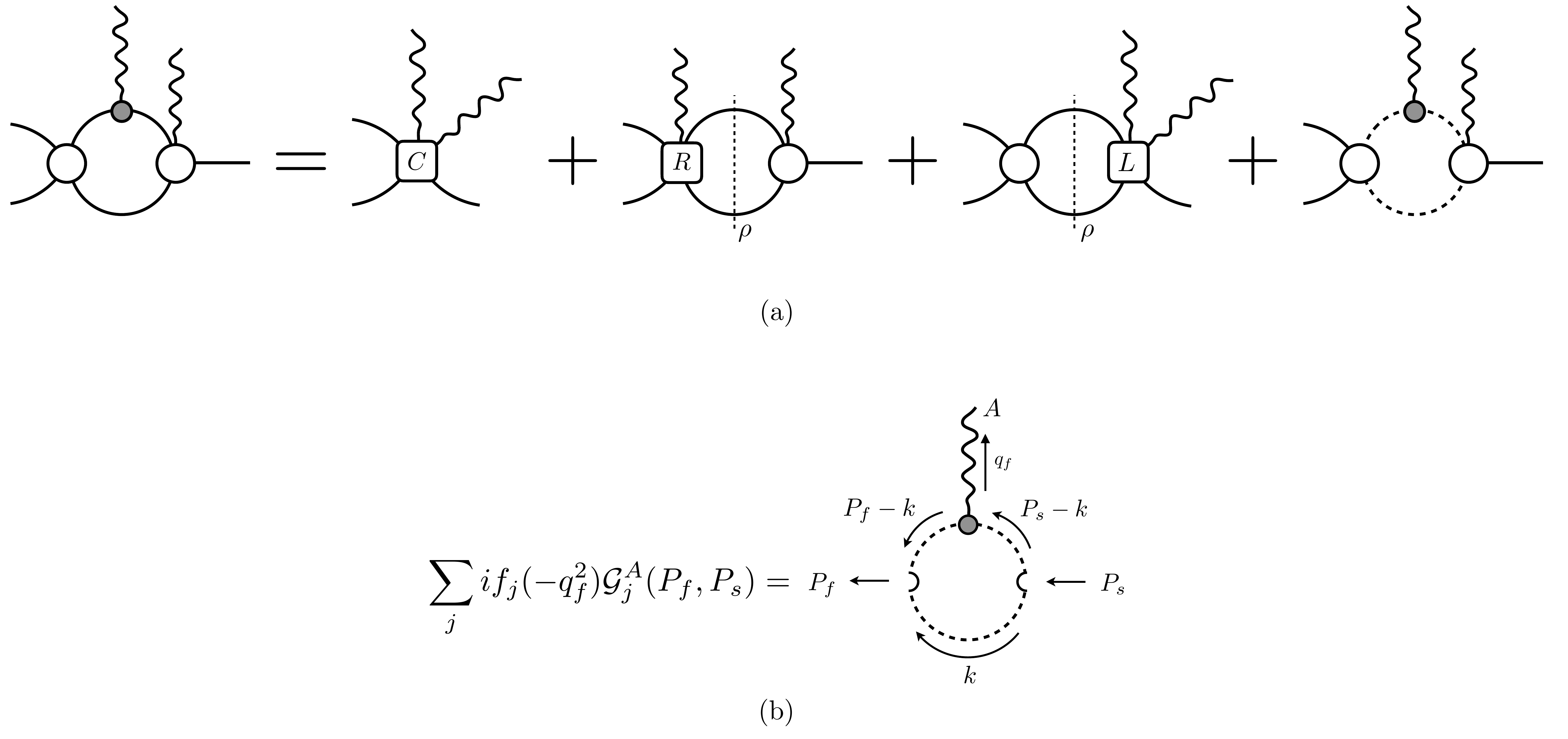}
\caption{ 
Shown in (a) is the decomposition of the triangle diagram into each of its on-shell pieces where the first three terms contain the smooth kernels $\Tb_{0|C|0}$, $\Wb_{0|R}$, and $\Tb_{L|0}$ respectively. The analytic form of this is shown in Eq.~\eqref{eq:triag_term1}. The final term of this decomposition is shown in more detail in (b) where the open semi circles on the left and right represent the modified spherical harmonics $\Yc_{\ell m_\ell}$.
}
\label{fig:triangle}
\end{center}
\end{figure}

Finally, we need to study the analytic structure arising from the triangle diagram in the last term of Eq.~\eqref{eq:C1B_dse}. To simplify the derivation we begin by replacing $\Mc$ and $\Hc$ with $\Kc_0$ and $\Hb_0$, respectively. Adding the rescattering contributions afterwards is straightforward.
This diagram was discussed in great detail in Ref.~\cite{Briceno:2020vgp}, and it was shown that this contribution can be decomposed into four pieces that feature different analytic behavior~
\begin{multline}
\label{eq:triag_term1}
\int
\frac{\diff^4k}{(2\pi)^4}
i\Kc_0(p',k) \Delta^{(2)}(P_{f},k) iw^{A}(k_f,k_s) i\Delta(k_s)  i\Hb_0^{B}(P_s,k;P_i)
= 
i\Tb^{AB}_{0|C|0}(P_f,p';P_i;q_f)
\\+
\sum_{\ell, m_\ell}i\Wb_{0|R,\ell m_\ell}^A(P_f,p';P_s) \,\rho\, i\Hb^B_{0,\ell m_\ell}(P_s,P_i)
+
\sum_{\ell, m_\ell}i\Kc_{0,\ell m_\ell}(p') \,\rho\, i\Tb^{AB}_{L|0,\ell m_\ell}(P_f,P_i,q_f)
\\+
\sum_{\ell', m_\ell'}\sum_{\ell, m_\ell}\Kc_{0,\ell' m_\ell}(p')\sum_j if_j(-q_f^2)\Gc^A_{j,\ell' m_\ell',\ell m_\ell}(P_f,P_s)
\Hb_{0,\ell m_\ell}^B(P_s,P_i)\,,
\end{multline}
where $\Gc$ is the triangle function defined in Eq.~\eqref{eq:G.function} which encodes the possible triangle singularities associated with all intermediate particles in Fig.~\ref{fig:triangle}(b) going on shell. The kernel $\Wb_{0|R}$ is the mirror of $\Wb_{L|0}$, i.e.\ it is a smooth function. The kernel $\Tb_{0|C|0}$ captures the remaining analytic behavior of the triangle diagram. 

It is worth commenting on the subscript of the $\Tb$ kernels. First, $\Tb_{0|C|0}$ denotes a kernel that is arising from a short distance contribution where a single-particle coupling to the current appears in the \emph{center} of the triangle diagram. In contrast to this, the kernel $\Tb_{L|0}$ arises from the one-body contribution being to the right of the triangle diagram, but to the \emph{left} of $\Hb_0$, and it is the same kernel that appeared in Eq.~\eqref{eq:T21.TL0def}. Finally, $\Tb_{0|R}$ appeared when the one-body contribution was to the \emph{right} of the $\Hb_0$ in Eq.~\eqref{eq:T21_1B_sinpart}. In Fig.~\ref{fig:triangle}(a) we illustrate the different contributions to the triangle diagram.

Equation~\eqref{eq:triag_term1} is, of course, one contribution to the last term in Eq.~\eqref{eq:C1B_dse}. This contribution will be dressed by an infinite number of terms with $s$-channel integrals of the form $\int \Kc_0 \Delta^{(2)}$ from the left, and in the intermediate state between the one-body current and $\Hb_0$. These integrals will result in further $\rho$ and triangle singularities.
It is straightforward to see that for the intermediate states we will need to consider triangle diagrams of the form $\int i\Kc_0 \cdot \Delta^{(2)} \cdot i w  \cdot i \Delta \cdot i \Kc_0$. Expressions of these terms can be obtained by replacing the kernels $\Tb\to \Wb$ and $\Hb_0\to \Kc_0$ into the previous equation. All the terms proportional to the triangle function $\Gc$ can be grouped and summed together to recover the $\Mc$ and $\Hc$ amplitudes. In short, this can be obtained from the last term in the previous equation using the following replacement,
\begin{align}
    \sum_{\ell', m_\ell'}\sum_{\ell, m_\ell}\Kc_{0,\ell' m_\ell}(p')\sum_j if_j(-q_f^2)\Gc^A_{j,\ell' m_\ell',\ell m_\ell}(P_f,P_s)
\Hb_{0,\ell m_\ell}^B(P_s,P_i)
&\nn\\
&\hspace{-5cm}\longrightarrow
\sum_{\ell', m_\ell'}\sum_{\ell, m_\ell}\Mc_{\ell'}(p')\sum_j if_j(-q_f^2)\Gc^A_{j,\ell' m_\ell',\ell m_\ell}(P_f,P_s)
\Hc_{ \ell m_\ell}^B(P_s,P_i).
    \label{eq:triangle_contribution}
\end{align}

We now turn our attention to the determination of terms that do not include the triangle function $\Gc$. At first we will only focus on the additional terms from intermediate two-body rescattering; the $s$-channel integrals from the final state interactions are straightforward to account for at the end by means of Eq.~\eqref{eq:App.FSIfin}. We start by considering terms that will dress $\Tb^{AB}_{0|C|0}$ to the right. These arise from the analytic piece of the triangle diagrams, i.e.\ $\Tb_{0|C|0}$, or terms containing $\Wb_{0|C|0}$. After grouping them they can be expressed as
\begin{multline}\label{eq:TCinfWcinf}
i\Tb^{AB}_{0|C|0}(P_f,p';P_i;q_f)
+
\xi\int
\frac{\diff^4k}{(2\pi)^4}
i\Wb^{A}_{0|C|0}(P_f,p';P_s,k)\Delta^{(2)}(P_{s},k)i\Hc^B(P_f,k;P_i)
\\
=i\Tb^{AB}_{0|C|\infty}(P_f,p';P_i;q_f)
+ \sum_{\ell, m_\ell}i\Wb^{A}_{0|C|\infty,\ell m_\ell}(P_f,p';P_s)\,\rho\,
i\Hc^B_{\ell m_\ell}(P_s,P_i)\,,
\end{multline}
where we have used Eq.~\eqref{eq:App.MSIsol} to project the kernels on shell, and defined the smooth kernel $\Tb_{0|C|j}^{AB}$ implicitly via the loop identity
\begin{multline}
    \xi\int
\frac{\diff^4k}{(2\pi)^4}
i\Wb^{A}_{0|C|j}(P_f,p';P_s,k)\Delta^{(2)}(P_{s},k)i\Hb^B_0(P_s,k;P_i)
\\=i\Tb^{AB}_{0|C|j+1}(P_f,p';P_i;q_f) +\sum_{\ell, m_\ell}
i\Wb^{A}_{0|C|j,\ell m_\ell}(P_f,p';P_s)
\,\rho \, i\Hb^B_{0,\ell m_\ell}(P_f,P_i)\,.
\label{eq:defT0Cj}
\end{multline}

These same kernels will later be dressed from the left from contributions of the form, 
\begin{multline}
    \xi\int
\frac{\diff^4k}{(2\pi)^4}
i\Kc_{0}(p',k)\Delta^{(2)}(P_{s},k)
i\Tb^{AB}_{j|C|j'}(P_s,k;P_i;q_f)
\\=i\Tb^{AB}_{j+1|C|j'}(P_f,p';P_i;q_f) +\sum_{\ell, m_\ell}
i\Kc_{0,\ell m_\ell}(p';P_s)
\,\rho \, i\Tb^{AB}_{j+1|C|j',\ell m_\ell}(P_f,P_i,q_f)\,.
\label{eq:defTjCj'}
\end{multline}

Next, the kernel $\Wb_{0|R}$ in the triangle diagram decomposition is always featured next to a phase space factor $\rho$, so that all terms to the right of them can be factored into the amplitude $\Hc$ and placed on-shell
\begin{multline}\label{eq:W0RH}
\sum_{\ell, m_\ell}i\Wb_{0|R,\ell m_\ell}^A(P_f,p';P_s) \,\rho\,\qty( i\Hb^B_{0,\ell m_\ell}(P_s,P_i)
+
\xi\int
\frac{\diff^4k}{(2\pi)^4}
i\Kc_{0,\ell m_\ell}(k)
\Delta^{(2)}(P_{f},k)
i\Hc^B(P_f,k;P_i))
\\=
\sum_{\ell ,m_\ell}i\Wb_{0|R,\ell m_\ell}^A(P_f,p';P_s) \,\rho\, i\Hc^B_{\ell m_\ell}(P_s,P_i)\,.
\end{multline}
The final state interactions can be taken into account in Eqs.~\eqref{eq:TCinfWcinf}~and~\eqref{eq:W0RH} by applying the result found in Eq.~\eqref{eq:App.FSIfin},
\begin{align}
i\Tb^{AB}_{0|C|\infty}
&\longrightarrow
i\Tb^{AB}_{\infty|C|\infty} 
+
\sum_{\ell,m_\ell}
i\Mc_{\ell}\,\rho\,i\Tb^{AB}_{\infty|C|\infty,\ell m_\ell} \,,\\
i\Wb^{A}_{0|C|\infty,\ell m_\ell}
&\longrightarrow
i\Wb_{\infty|C|\infty,\ell m_\ell}^A
+
\sum_{\ell',m_\ell'}
i\Mc_{\ell'}\,\rho\,i\Wb_{\infty|C|\infty,\ell' m_\ell',\ell m_\ell}^A\,,\\
i\Wb_{0|R,\ell m_\ell}^A 
&\longrightarrow
i\Wb_{\infty|R,\ell m_\ell}^A
+
\sum_{\ell',m_\ell'}
i\Mc_{\ell'}\,\rho\,i\Wb_{\infty|R,\ell' m_\ell',\ell m_\ell}^A\,.
\end{align}

Finally, all the terms with one-body contributions to the left of the kernel after splitting the analytic behavior of the triangle diagrams feature either $\Tb_{L|0}$ or $\Wb_{L|0}$. These terms will appear within a sum over partial waves times a factor of $i\Kc_{0,\ell m_\ell}(p') \,\rho$, e.g.\ the third term on the right hand side of Eq.~\eqref{eq:triag_term1}. Once final state interactions are taken into account, this factor becomes $i\Mc_{\ell}(p') \,\rho$. Leaving this common factor implied, the analytic structure of the terms with \emph{left} kernels can be found, with the aide of Eq.~\eqref{eq:App.MSIsol}, to be equal to
\begin{multline}\label{eq:TLinfWLinf}
 i\Tb^{AB}_{L|0,\ell m_\ell}(P_f,P_i,q_f)
+
\xi\int
\frac{\diff^4k}{(2\pi)^4}
i\Wb^{A}_{L|0}(P_f,p';P_s,k)\Delta^{(2)}(P_{s},k)i\Hc^B(P_f,k;P_i)
\\
=
 i\Tb^{AB}_{L|\infty,\ell m_\ell}(P_f,P_i,q_f)
+ \sum_{\ell', m_\ell'}i\Wb^{A}_{L|\infty,\ell m_\ell,\ell' m_\ell'}(P_f,P_s)\,\rho\,
i\Hc^B_{\ell' m_\ell'}(P_s,P_i)\,,
\end{multline}
where $\Tb_{L|j}^{AB}$ is defined via the loop identity that results from replacing the subscript $0|C\to L$ in the $\Wb$ and $\Tb$ kernels of Eq.~\eqref{eq:defT0Cj}.

The on-shell projection of the $\Tc_{1\rm{B}}$ projection can be found by adding together Eqs.~\eqref{eq:C1B_dse_term1_vf}, \eqref{eq:C1B_dse_term2_vf}, \eqref{eq:triag_term1} along with the rescattering contributions for Eq.~\eqref{eq:triag_term1},
\begin{multline}
    i\Tc^{AB}_{21,1B}(P_f,p';P_i;q_f)
    = i\Hc^{A}_{ \mathrm{on}}(P_{f},p';P_{s}) iD({s}) iw^{B}_{ \mathrm{on}}(P_{s},P_{i})
    + iw^{A}_{ \mathrm{on}}(p'_f,p'_s) iD(p'^2_s) i\overline{\Hc}^B(P_s,p';P_i)
    %%%%%%%%%%%%%%%%%%%%%%%%%%%%%
\\
+\sum_{\ell, m_\ell} \sqrt{4\pi}Y_{\ell m_\ell}(\hat{\mathbf{p}}')
[1+ i\Mc_{\ell}(s_f)\,\rho]\Big(i\Tb_{1B,\ell m_\ell}^{AB}(P_f,P_i,q_f)
    +\sum_{\ell', m_\ell'}i\Wb^A_{1B,\ell m_\ell,\ell'm_\ell'}(P_f,P_s)\,\rho\, i\Hc^B_{\ell' m_\ell'}(P_s,P_i)\Big)
    \\+\sum_{\ell, m_\ell} \sqrt{4\pi}Y_{\ell m_\ell}(\hat{\mathbf{p}}')
    \Mc_\ell(s_f)\sum_{\ell'm_\ell'}\sum_{j}if_j(-q_f^2)\Gc^A_{j,\ell m_\ell,\ell'm_\ell'}(P_f,P_s)\Hc^B_{\ell'm_\ell'}(P_s,P_i)\,,
\label{eq:C1B_dse_vf}
\end{multline}
where we have added the smooth contributions into a single smooth function 
\begin{equation}
    \Tb^{AB}_{1\rm{B}} = \Tb^{AB}_{\infty|C|\infty} +  \Tb^{AB}_{L|\infty} 
    +\Tb^{AB}_{\infty|R}\,, 
\end{equation}
and $\Wb^A_{1\rm{B}}$ is the same as defined in \cite{Briceno:2020vgp}, 
\begin{align}
    \Wb_{1\rm{B}}^{A} = \Wb_{\infty|C|\infty}^{A} + \Wb_{L|\infty}^{A} + \Wb_{\infty|R}^{A} \, .
\end{align}

%%%%%%%%%%%%%%%%%%%%%%%%%%%%%%%%%%%%%%%%%%%
%  Subsubsection: Full on-shell result
%%%%%%%%%%%%%%%%%%%%%%%%%%%%%%%%%%%%%%%%%%%
\subsubsection{Full on-shell result for $\Tc_{21}$}

We began the derivation of the on-shell amplitude $\Tc_{21}$ by separating it into two terms. The first, labeled $\Tc_{21,\cancel{1\rm{B}}}$, is defined to include all possible diagrams that do not include any one-body contribution, and it satisfies 
Eq.~\eqref{eq:Cno1B_dse}.
The second, labeled $\Tc_{21,1B}$, includes all contributions where a single-particle can couple to the external current directly. This set of diagrams satisfy Eq.~\eqref{eq:C1B_dse}. By projecting all possible intermediate states that may go on-shell, we showed these two terms could be written in terms of purely on-shell functions as Eqs.~\eqref{eq:Cno1B_dse_vf} and \eqref{eq:C1B_dse_vf}, respectively. 

Throughout the derivation, we have only included the contributions from the direct channel for the current insertions. In particular, we have assumed that the momentum of current $B$ is inserted into the initial state, and that the $A$ current takes momentum from the intermediate state. As discussed in Sec.~\ref{sec:1Jto1J}, the exchange contributions can be obtained by first swapping the $A$ and $B$ labels, followed by changing $q_f\to -q_i$, which results in $s\to u$. 

Adding Eqs.~\eqref{eq:Cno1B_dse_vf} and \eqref{eq:C1B_dse_vf} as well as their $u$-channel contributions, we arrive at the final expression for $\Tc_{21}$,
\begin{multline}
    i\Tc_{21}(P_f,p';P_i;q_f)
   = 
   i\Hc^{A}_{ \mathrm{on}}(P_{f},p';P_{s}) iD({s}) iw^{B}_{\mathrm{on}}(P_{s},P_{i})
   +
   i\Hc^{B}_{ \mathrm{on}}(P_{f},p';P_{u}) iD({u}) iw^{A}_{\mathrm{on}}(P_{u},P_{i})
    \\
    + iw^A_{\mathrm{on}}(p'_f,p'_s) iD(p'^2_s) i\overline{\Hc}^B(P_s,p';P_i)
     + iw^B_{\mathrm{on}}(p'_f,p'_u) iD(p'^2_u) i\overline{\Hc}^A(P_u,p';P_i)
    +i\Tc_{21,\df} (P_f,p';P_i;q_f)
    \,,
    \label{eq:T21_vf}
\end{multline}
where, 
\begin{align}
    i\Tc_{21,\df}(P_f,p';P_i;q_f)
    \equiv
    i\Tc_{21,\df}^{AB}(P_f,p';P_i;q_f)
    +i\Tc_{21,\df}^{BA}(P_f,p';P_i;-q_i).
\end{align}
As was the case with $\mathcal{W}$ and $\Tc_{11}$, we have isolated the simple pole singularities and defined the remainder of the amplitude as $\Tc_{21,\df}$. 
We now proceed to give a compact expression for $\Tc_{21,\df}^{AB}$ in terms of on-shell physical quantities. After doing so, we will be able to add the exchange contribution. 
The divergence-free terms appearing in $\Tc_{21,\df}^{AB}$ can be written as
\begin{multline}
    i\Tc_{21,\df,\ell m_\ell}^{AB}(P_f,P_i,q_f)
    = 
[1+ i\Mc_{\ell}(s_f)\rho]\Big(i\Tb_{\ell m_\ell}^{AB}(P_f,P_i,q_f)
    +\sum_{\ell', m_\ell'}i\Wb^A_{\ell m_\ell,\ell'm_\ell'}(P_f,P_s)\rho i\Hc^B_{\ell' m_\ell'}(P_s,P_i)\Big)
   \\ +
    \Mc_\ell(s_f)\sum_{\ell'm_\ell'}\sum_{j}if_j(-q_f^2)\Gc^A_{j,\ell m_\ell,\ell'm_\ell'}(P_f,P_s)\Hc^B_{\ell'm_\ell'}(P_s,P_i)\,,
    \label{eq:T21df}
\end{multline}
where we have added together $\Wb_{\cancel{1\rm{B}}}$ and $\Wb_{1\rm{B}}$ and $\Tb_{\cancel{1\rm{B}}}$ and $\Tb_{1\rm{B}}$ into single kernels $\Wb$ and $\Tb$ respectively.  

As discussed in Ref.~\cite{Briceno:2020vgp}, $\Wb$ has $K$-matrix poles associated with the rescattering of initial and final states. A similar behavior is followed by $\Tb$, except that for this function the poles are associated with the rescattering of final and \emph{intermediate} two-particle states. We make the possible unphysical poles in the $K$-matrix explicit, by parameterizing these functions as
\begin{align}
    \Tb_{\ell m_\ell}^{AB}(P_f,P_i,q_f) &= \Kc_\ell(s_f)\Bc^{AB}_{21,\ell m_\ell}(P_f,P_i,q_f)
    +\Kc_\ell(s_f)\sum_{\ell', m_\ell'}\Ac^A_{22,\ell m_\ell,\ell'm_\ell'}(P_f,P_s)\Kc_{\ell'}(s)\Ac^B_{21,\ell'm_\ell'}(P_s,P_i)\,,\\
    \Wb_{\ell m_\ell,\ell' m_\ell'}^{A}(P_f,P_s)&=
    \Kc_\ell(s_f)\Ac^A_{22,\ell m_\ell,\ell'm_\ell'}(P_f,P_s)\Kc_{\ell'}(s),
\end{align}
where the $\Bc_{21}$, $\Ac_{21}$, and  $\Ac_{22}$ are smooth functions and the second equality is identical to the one used in Ref.~\cite{Briceno:2020vgp}. It is important to emphasize that $\Ac_{22}$ and $\Ac_{21}$ are the same functions that appear in the definition of $\Hc$ and $\Wc_\df$, given in Eqs.~\eqref{eq:H_on-shell} and \eqref{eq:Wdf_def}, respectively. 

Using these parameterizations, we can rewrite Eq.~\eqref{eq:T21df} in terms of quantities that do not depend on the unphysical $K$-matrix poles,
\begin{multline}
    \label{eq:T21df_v2}
    i\Tc_{21,\df,\ell m_\ell}^{AB}(P_f,P_i,q_f)
    = 
    i\Mc_{\ell}(s_f)\Bc^{AB}_{21,\ell m_\ell}(P_f,P_i,q_f)
    \\+
    \Mc_{\ell}(s_f)
    \sum_{\ell', m_\ell'}\Big(i\Ac^{A}_{22,\ell m_\ell}(P_f,P_s)
    +\sum_{j}if_j(-q_f^2)\Gc^A_{j,\ell m_\ell,\ell'm_\ell'}(P_f,P_s)
    \Big)
    \Mc_{\ell'}(s)\Ac^B_{21,\ell'm_\ell'}(P_s,P_i)\,.
\end{multline}
This can be further simplified by recognizing that the term inside of the parenthesis is exactly equal to $\Wc_\df$, given in Eq.~\eqref{eq:Wdf_def}. Making this replacement and adding the exchange diagrams, we arrive at our final expression for $\Tc_{21,\df}$,
\begin{align}
    i\Tc_{21,\df,\ell m_\ell}(P_f,P_i,q_f)
    &= 
    i\Mc_{\ell}(s_f)\Bc_{21,\ell m_\ell}(P_f,P_i,q_f)
\nn\\
    &\hspace{-1.5cm}
    +
    \sum_{\ell', m_\ell'}
    i\Wc^A_{\df,\ell m_\ell,\ell'm_\ell'}(P_f,P_s)\Ac^B_{21,\ell'm_\ell'}(P_s,P_i)\,
    +
    \sum_{\ell', m_\ell'}
    i\Wc^B_{\df,\ell m_\ell,\ell'm_\ell'}(P_f,P_u)\Ac^A_{21,\ell'm_\ell'}(P_u,P_i)\,,
    \label{eq:T21df_vf}
\end{align}
where $\Bc_{21}\equiv\Bc^{AB}_{21}+\Bc^{BA}_{21}$. This is the final and main result of this section.

%%%%%%%%%%%%%%%%%%%%%%%%%%%%%%%%%%%%%%%%%%%
%  Subsection: Generalization
%%%%%%%%%%%%%%%%%%%%%%%%%%%%%%%%%%%%%%%%%%%
\subsection{Generalization to multiple channels and arbitrary masses} \label{sec:general_case}

In the derivation above we made a set of simplifying assumptions, which we proceed to lift here. In general, one needs to consider the possibility that: (a) the particles appearing inside the loops may have different masses, (b) the current can couple to both intermediate particles, (c) any number of two-particle channels may be kinematically open, and (d) the current may couple different single-particle states, i.e. $1+\Jc \to 1'$. 

Allowing for different masses is straightforward. The role of the masses is encoded in the $\rho$ and $\Gc$ kinematic functions, which have been written in Eq.~\eqref{eq:ps} and \eqref{eq:G.function} for arbitrary masses. 

If the current can couple to both external single-particle states, the divergent part of $\Tc_{21}$, shown on the first four terms of Eq.~\eqref{eq:T21_vf}, acquires additional pole terms associated with these couplings. If both of the intermediate single-particle states also couple to the current, there will be an additional triangle function contribution to the $\Wc_\df$, but Eq.~\eqref{eq:T21df_vf} as is written now will remain unchanged. 

One can accommodate any number of intermediate two-particle channels by upgrading the on-shell kernels into either vectors or matrices in channel space~\cite{Briceno:2012yi, Hansen:2012tf, Briceno:2014oea}. More explicitly, if we suppress the angular momentum indices of the amplitudes, we can rewrite $\Tc_{21,\df}$ for arbitrary channels as, 
\begin{align}
    i\Tc_{21,\df, a }(P_f,P_i,q_f)
    &= 
    i\Mc_{ab}(s_f)\Bc_{21,b}(P_f,P_i,q_f)
\nn\\
    &\hspace{-1.5cm}
    +
    i\Wc^A_{\df,ab}(P_f,P_s)\Ac^B_{21,b}(P_s,P_i)\,
    +
    i\Wc^B_{\df,ab}(P_f,P_u)\Ac^A_{21,b}(P_u,P_i)\,,
    \label{eq:T21df_general}
\end{align}
where the ``$b$" index runs over the possible intermediate channels. We can then label the masses of the two particles present in the $a$-th channel as $m_{a1}$ and $m_{a2}$. The single particle form factors would also get indices associated with the particle type, e.g.\ $f_j\to f_{j,a1}$. 

Finally, the kinematic factor associated with the Lorentz decomposition of the current, would also have to acquire an index associated with the particle it is coupling, e.g. $K_{j}\to  K_{j, a1}$. To understand this, it is useful to consider the case where the current is a vector and the initial and final state are the same particle with mass $m_{a1}$. Considering the triangle diagram, Fig.~\ref{fig:triangle} (b), and giving the spectator particle a momentum $k$, the on-shell projected one-body transition would have the standard Lorentz composition in terms of a single form factor, 
\begin{align}
    iw^{\mu}_{\mathrm{on},a1}(k_f,k_i)
    &= (P_f+P_{i}-2k)^\mu \big|_{k^2 = m^2_{a2}}f(Q^2) \nn\\
    &\equiv K^\mu_{a1}(k_f,k_i)  f(Q^2) .
\end{align}
From this example we see two things. First, the mass of the particles appears as a constraint in $k^\mu$. Second, this depends on the mass of the spectator, not the mass of the particle that it is coupling to. Despite this, we choose to label the kinematic factor with the label associated with the particle that couples to the current. 

Equation~\eqref{eq:T21df_general} also accommodates the case where the current couples different single particle states. The only subtlety that arises in this case, which was discussed in Ref.~\cite{Briceno:2020vgp}, is that the $\Gc$ function will not be diagonal over channel space.~\footnote{We do not provide explicit expressions for the $\Gc$ function for this case, but they can be found in Ref.~\cite{Briceno:2020vgp}.}

%%%%%%%%%%%%%%%%%%%%%%%%%%%%%%%%%%%%%%%%%%%
%  Section: Conclusion
%%%%%%%%%%%%%%%%%%%%%%%%%%%%%%%%%%%%%%%%%%%
\section{Conclusion}\label{sec:conclusion}

We have presented an on-shell representation for transition amplitudes from a single-hadron state to a two-hadron state induced by two external currents. The framework is constructed in a model independent fashion by summing to all orders in the strong interaction and building off the previously determined on-shell relations for $2\to 2$, $1+\Jc\to 2$, and $2+\Jc\to2$ amplitudes~\cite{Briceno:2020vgp}. The result presented here is valid in the kinematic range below three-particle threshold where any number of two-hadron channels may be open, for currents with arbitrary Lorentz structure, and spinless hadrons. 
The resulting amplitudes contain the usual threshold branch cuts in the final and intermediate two-particle energies, as well as logarithmic singularities arising from the triangle diagrams in both the direct and exchange channels. In order to describe reactions of spinful particles further work is necessary to understand the analytic behavior of each kernel and loop as a matrix in spin space, particularly the triangle diagram which in general will be a dense matrix in this space.

We showed that analytically continuing to the poles in the unphysical Riemann sheet in the final state energy allows for access to the Compton-like amplitudes coupling a single particle to resonant states. Furthermore, by taking the initial particle together with one of the currents, we showed that the analytic continuation of their total energy recovers the definition of the elastic resonance form factors as found in previous work~\cite{Briceno:2020vgp}, showing consistency of this formalism. For the case of conserved vector currents, we showed the Ward-Takahashi identity places constraints on the amplitude such that when one of the current momenta vanishes the short distance piece may be given in terms of the subprocess amplitudes.

In the context of precision tests of the Standard Model, this formalism can contribute to the determination of hadronic light-by-light amplitudes required to reduce the theoretical uncertainty on the anomalous magnetic moment of the muon. Similarly, it supplements the constraints imposed by $\chi$PT in low energy Standard Model observables such as the rare Kaon decay $K\to\pi\pi\gamma^{\star}$~\cite{NA482:2010iwd,NA482:2018gxf,Cappiello:2017ilv}. This formalism also allows access to better understanding of the hadronic structure of resonances through analysis of $\gamma^{\star}\gamma^{\star}\to\pi\pi,K\overline{K},\ldots$ transition amplitudes as a way to search for glueballs in the isoscalar sector. When combined with a future finite-volume framework, we envision this formalism being a useful tool for lattice QCD calculations of the aforementioned processes. 

Finally, we also emphasize that the Compton-amplitude reviewed in this work acts as a stepping stone to understanding the more complicated two-current amplitude with two hadrons in the initial \emph{and} final state. Such amplitudes would be required for studying Compton scattering of two-body shallow bound states and resonances as well as neutrino-less double beta decay. In general this amplitude will be a function of the allowed subprocesses, thus having a rigorous understanding of them is key.
 
In summary, the presented formalism is useful for phenomenological studies probing the Standard Model as well as for accessing the structure of resonant hadronic states. Some of these cases may be studied via the use of lattice QCD, pending a future finite-volume framework, where this formalism would be immediately relevant as it would allow for the extraction of physical observables.

%%%%%%%%%%%%%%%%%%%%%%%%%%%%%%%%%%%%%%%%%%%
%  Section: Acknowledgements
%%%%%%%%%%%%%%%%%%%%%%%%%%%%%%%%%%%%%%%%%%%
\section{Acknowledgements}
The authors would like to thank J.~Dudek and M.~Hansen for useful comments on the manuscript, as well as A.~Rodas for useful discussions.
RAB and AWJ acknowledges support from U.S. Department of Energy contract DE-AC05-06OR23177, under which Jefferson Science Associates, LLC, manages and operates Jefferson Lab.
FGO  acknowledges support from the U.S.\ Department of Energy contract DE-SC0018416 at William \& Mary and the JSA/JLab Graduate Fellowship Program.
KHS, AWJ, and RAB acknowledge support of the USDOE Early Career award, contract DE-SC0019229.
KHS acknowledges support by the U.S. Department of Energy, Office of Science Graduate Student Research (SCGSR) program. The SCGSR program is administered by the Oak Ridge Institute for Science and Education (ORISE) for the DOE. ORISE is managed by ORAU under contract number DE-SC0014664. All opinions expressed in this paper are the author’s and do not necessarily reflect the policies and views of DOE, ORAU, or ORISE.

\appendix

%%%%%%%%%%%%%%%%%%%%%%%%%%%%%%%%%%%%%%%%%%%
%  Section: Lorentz decomposition of Compton scattering
%%%%%%%%%%%%%%%%%%%%%%%%%%%%%%%%%%%%%%%%%%%
\section{Lorentz decomposition of Compton scattering for low-energy photons}
\label{sec:App.Loren}

The behavior of the amplitude $1+\Jc^\nu\to1+\Jc^\mu$ in the limit of vanishing initial and final photon energy is determined by the analytic structure of the amplitude $\Tc^{\mu\nu}_{11}$ and gauge invariance. As derived in Sec.~\ref{sec:1Jto1J} and shown in Eq.~\eqref{eq:T11_on-shell}, within the kinematic region of interest this amplitude contains an analytic piece and two pole contributions. For energies well below the two-particle threshold, we can compactly rewrite the amplitude in the form %
\begin{equation}
    i\Tc^{\mu\nu}_{11}(P_f,P_i,q_f)
    = i\beta_{11}^{\mu\nu}(P_f,P_i,q_f)
    +
    iw_{\mathrm{on}}^{\mu}(P_f,P_s) iD(s) iw_{\mathrm{on}}^{\nu}(P_s,P_i)
    +
    iw_{\mathrm{on}}^{\nu}(P_f,P_u) iD(u) iw_{\mathrm{on}}^{\mu}(P_u,P_i)\,,
\end{equation}
where both $s$- and $u$-channel contributions are explicitly shown, and the remainder, which absorbs the smooth function $\Bc_{11}$ and the two-body contribution that has been analytically continued well below threshold, is expressed in a single smooth function $\beta_{11}^{\mu\nu}$. Gauge invariance on the other hand imposes that
\begin{equation}\label{eq:app.Lor.gaug}
    q_{f\mu}\Tc_{11}^{\mu\nu}(P_f,P_i,q_f)=q_{i\nu}\Tc_{11}^{\mu\nu}(P_f,P_i,q_f) = 0\,.
\end{equation}
These equations impose constraints on the analytic piece $\beta_{11}$ in terms of the pole pieces. To show this we first contract the final photon momentum with the amplitude
\begin{equation}\label{eq:app.wardrel}
 q_{f\mu}i\beta_{11}^{\mu\nu}(P_f,P_i,q_f) = -
 q_{f\mu}\bigg( iw_{\mathrm{on}}^{\mu}(P_f,P_s) iD(s) iw_{\mathrm{on}}^{\nu}(P_s,P_i)
    +
    iw_{\mathrm{on}}^{\nu}(P_f,P_u) iD(u) iw_{\mathrm{on}}^{\mu}(P_u,P_i) \bigg)\,.
\end{equation}
Assuming the external particle is spinless we can simplify the right-hand side of this expression with the help of the Lorentz decomposition of the three point function shown in Eq.~\eqref{eq:WT.wonLD},
\begin{align}
    q_{f\mu}i\beta_{11}^{\mu\nu}(P_f,P_i,q_f) &= 
    f(Q_f^2)\bigg(iw_{\mathrm{on}}^{\nu}(P_s,P_i) - iw_{\mathrm{on}}^{\nu}(P_f,P_u)
    \bigg)\,,\\
    q_{i\nu}i\beta_{11}^{\mu\nu}(P_f,P_i,q_f) &= 
     f(Q_i^2)
     \bigg(iw_{\mathrm{on}}^{\mu}(P_f,P_s)- iw_{\mathrm{on}}^{\mu}(P_u,P_i)
     \bigg)\,,
\end{align}
where the second line is the result of applying the same procedure to Eq.~\eqref{eq:app.Lor.gaug}.

To obtain the behavior at vanishing photon momenta of the Compton scattering amplitude, we first make the initial and final external momenta equal $P_i=P_f=P$, which in turn enforces that $q_f=q_i=q$. By using the explicit Lorentz decomposition of the single particle transition in Eq.~\ref{eq:WT.wonLD}, the Ward identities simplify to
\begin{align}
     q_{\mu}i\beta_{11}^{\mu\nu}(P,P,q) &=2iq^\nu f(Q^2)^2\,,\\
    q_{\nu}i\beta_{11}^{\mu\nu}(P,P,q) &=2iq^\mu f(Q^2)^2\,.
\end{align}
Finally, we expand around the limit of vanishing photon momentum, and by equating the first term of the series in each side of both equations, we obtain the result
\begin{equation}
    i\beta_{11}^{\mu\nu}(P,P,0) = 2ig^{\mu\nu}f(0)^2\,.
\end{equation}

To obtain the behavior of the full amplitude at vanishing photon momentum we again begin by making the initial and final external particle momentum equal
\begin{align}
    i\Tc_{11}^{\mu\nu}(P,P,q)&=
    i\beta_{11}^{\mu\nu}(P,P,q)+
    f(Q^2)^2( i(2P+q)^\mu iD(s) i(2P+q)^\nu 
    +i(2P-q)^\nu iD(u) i(2P-q)^\mu)\\
&= i\beta_{11}^{\mu\nu}(P,P,q)
-if(Q^2)^2(
(4P^\mu P^\nu+q^\mu q^\nu)(D(s)+D(u)) +2(P^\mu q^\nu+P^\nu q^\mu)(D(s)-D(u)
)\,.
\end{align}
To take the limit of vanishing photon momentum we choose the rest frame of the external particle such that the propagators can be written as
\begin{align}
    D(s)^{-1} &= 2mq_0+q^2+i\epsilon\,,\\
    D(u)^{-1} &= -2mq_0+q^2+i\epsilon\,,
\end{align}
where $m$ is the mass of the external particle. Since we are interested in the case of real Compton scattering, $q^2=0$. We will distinguish this process with the notation $i\Tc^{\mu\nu}_{11,\mathrm {R}}$.
In this case, the propagators have the behavior
\begin{align}
\eval{D(s)+D(u)}_{q^2=0}= 0\,,\quad \eval{D(s)-D(u)}_{q^2=0} = \frac{1}{mq_0}\,.
\end{align}
Then the leading order behavior of the Compton amplitude is
\begin{equation}
i\Tc^{\mu\nu}_{11,\mathrm {R}}(P,P,q) = 2if(0)^2\qty(g^{\mu\nu}- (P^\mu q^\nu+P^\nu q^\mu)\frac{1}{mq_0}) +\mathcal{O}(q_0^2)\,.
\end{equation}
In this case, the amplitude involves real photons, so it needs to be contracted with the photon wave-functions $\epsilon(q)_{\mu}'$ and $\epsilon(q)_\nu$,
\begin{equation}
\epsilon_\mu(q)^{\prime*}\epsilon_\nu(q) i\Tc^{\mu\nu}_{11,\mathrm {R}}(P,P,q)= 2if(0)^2\epsilon(q)^{\prime*}\cdot \epsilon(q)  +\mathcal{O}(q_0^2)\,,
\end{equation}
and in the limit of zero photon energy we recover the Thomson amplitude,
\begin{equation}
\epsilon_\mu(0)^{\prime*}\epsilon_\nu(0) i\Tc^{\mu\nu}_{11,\mathrm {R}}(P,P,0) = 2if(0)^2{\boldsymbol \epsilon}^{\prime*}\cdot {\boldsymbol \epsilon}\,,
\end{equation}
where $\boldsymbol \epsilon^{(\prime)}$ is the polarization vector of the initial (final) photon. For the case of virtual photons we refer the reader to the discussion made in \cite{Drechsel:2002ar}.

%%%%%%%%%%%%%%%%%%%%%%%%%%%%%%%%%%%%%%%%%%%
%  Section: On-shell projection of transition amplitudes 
%%%%%%%%%%%%%%%%%%%%%%%%%%%%%%%%%%%%%%%%%%%
\section{On-shell projection of single-particle states}
\label{sec:App.osproj}

This appendix describes our prescription to perform the decomposition of the off-shell transition kernels $w$ and $\Hc$. The prescription to decompose the kernel $w$ whenever it appears within a two-particle state is given in App.~A of Ref.~\cite{Briceno:2020vgp}.
Here we are interested in the case where one of the external legs of the kernel $w$ or $\Hc$ is an intermediate single-particle state and may be off-shell. We find that the decomposition of Ref.~\cite{Briceno:2020vgp} for $w$ is also useful in this case, and use a similar strategy to decompose $\Hc$. This prescription is used in Eqs.~\eqref{eq:1Jto1.wDw_proj}~and~\eqref{eq:T21_1B_sinpart} of the main derivation. Within this appendix we assume all stable hadrons have mass $m$, the generalization to different masses being a straightforward exercise.

We begin with Eq.~\eqref{eq:1Jto1.wDw_proj}, where the three point function $w$ appears on both sides of the single-particle intermediate state
\begin{equation}\label{eq:AProj.wDw}
   \Cc_0(P_s) = iw^A(P_f,P_s)\cdot i\Delta(P_s)\cdot iw^B(P_s,P_i)\,,
\end{equation}
where $\Delta(P_s)$ is the fully-dressed single particle propagator as described in the main text. We make $\Cc_0$ a function of $P_s$ to emphasize that we are interested in the singularities arising in the kinematic variable $P_s^2$. A common choice to split the kinematic and dynamic behavior in transition amplitudes is to perform a Lorentz decomposition into form factors and kinematic tensors. For a given Lorentz structure of the external current in $w$ there will be a finite set of linearly independent tensors, so that for arbitrary values of the external momenta we have
\begin{equation}\label{eq:AProj.woffLD}
iw^A(p',p)= \sum_j K^A_j(p',p) if_j(Q^2,p^{\prime2},p^2)\,,
\end{equation}
where the form factors $f_j$ being Lorentz invariant can only depend on the Lorentz scalars $p^{\prime2},$ $p^2,$ and $Q^2=-(p^{\prime}-p)^2$.
The on-shell counterpart of this decomposition is given by Eq.~\eqref{eq:w_on}, from which we can recognize that the on-shell form factors are simply given by
\begin{equation}
    if_j(Q^2)=if_j(Q^2,m^2,m^2)\,.
\end{equation}

The generally off-shell form factor, can be written in terms of the partially and fully on-shell form factors using the $\delta$ operator used in Ref.~\cite{Briceno:2020vgp} and references therein,
\begin{equation}\label{eq:APRoj.ifexp}
    if_j(Q^2,p^{\prime2},p^2)
    = if_j(Q^2) + \delta[if_j(Q^2,p^{\prime2})] +
    [if_j(Q^2,p^{2})]\delta + \delta[if_j(Q^2,p^{\prime2},p^2)]\delta\,.
\end{equation}
In the limit that the initial (final) state is placed on-shell $f_j\delta$ ($\delta f_j$) is defined to vanish.

In the case of Eq.~\eqref{eq:AProj.wDw}, since $P_f$ and $P_i$ correspond to the momentum of the external final and initial state, they eventually will be placed on-shell. Assuming that the external particles have been placed on-shell, and substituting Eq.~\eqref{eq:APRoj.ifexp} into Eq.~\eqref{eq:AProj.woffLD} we obtain
\begin{align}
   \Cc_0(P_s)
    &= \sum_{j,l} K_j^A(P_f,P_s) if_j(-q_f^2) iD(P_s^{2})  K_l^B(P_f,P_s)if_l(-q_i^2) + \delta\Cc_0(P_s)\\
    &= iw^A_{\rm on}(P_f,P_s) iD(P_s^{2})  iw^B_{\rm on}(P_s,P_i) + \delta\Cc_0(P_s)\,,
    \label{eq:AProj.C0onshellexp}
\end{align}
where $q_{f(i)}=P_s-P_{f(i)}$ is as defined in the main text, and we have used the definition of $w_{\rm on}$ in Eq.~\eqref{eq:w_on}, where the kinematic pieces are evaluated with the momenta $P_s$, which is not necessarily on-shell. The simple pole piece of the propagator $D(P_s^{2})$ is given in Eq.~\eqref{eq:D_prop}.
From this we can conclude that the function $\delta\Cc(P_s)$ is smooth in the kinematic region of interest, this corresponds to the kernel $\Tb_{11,\alpha}$ of the main text.
Any chosen prescription should not change the singularity content of the amplitude. In this case, that means the residue of the one-particle pole in $\Cc_0$ has to be independent of the prescription. This boils down to ensuring that when the external legs are placed on-shell we recover the physical transition amplitude
\begin{equation}\label{eq:AProj.wonreq}
    \eval{iw^A_{\rm on}(p',p)}_{p^{\prime2}=p^2=m^2} = \mel{p'}{\Jc^A(0)}{p}\,.
\end{equation}

The second case of interest is the one appearing in Eq.~\eqref{eq:T21_1B_sinpart}.
However, before we discuss this on-shell expansion, we need to describe the form factor decomposition of a generic on-shell $1+\Jc\to 2$ amplitude. In particular we will focus on the case where the final two-particle state has been partial-wave projected to a definite angular momentum $\ell$ and projection in the $z$-axis $m_\ell$. In this case, the amplitude can be decomposed into form factors as follows
\begin{equation}\label{eq:AProj.HLD}
    \mel{P_f,\ell m_\ell}{\Jc^A(0)}{P_i} =
    \sum_j K^A_{j,\ell  m_\ell}(P_f,P_i) ih_{j,\ell}(s_f,Q^2)\,,
\end{equation}
where $Q^2 = -(P_f-P_i)^2$, and the functions $K_{j,\ell  m_\ell}$ are the corresponding kinematic factors that have to reproduce the Lorentz structure of the current. These differ from the $K_j$ in the decomposition of $w$, in that they also have to reproduce the non-trivial dependence on the angular momentum of the final state. Finally, the functions $h_{j,\ell}$ are energy-dependent transition form factors, which are Lorentz scalars irrespective of the nature of the current. The rotational properties of a given partial wave are contained within the kinematic factors, which means that the form factors $h_j$ are rotationally invariant and cannot depend on the azimutal component of the angular momentum $m_\ell$.

As summarized in Eq.~\eqref{eq:H_on-shell}, the transition amplitudes can be written in terms of $\Mc$ and the real-valued function $\Ac_{21}$. Given that $\Mc$ is independent of the Lorentz structure of the current, Eq.~\eqref{eq:AProj.HLD} provides the Lorentz decomposition of $\Ac_{21}$
\begin{equation}\label{eq:AProj.ALD}
    \Ac^A_{21,\ell m_\ell}(P_f,P_i) = \sum_j K^A_{j,\ell  m_\ell}(P_f,P_i) a_{j,\ell}(s_f,Q^2)\,.
\end{equation}
Assuming the $K_{j,\ell m_\ell}$ tensors have been defined without any spurious kinematic singularities, the $a_{j,\ell m_\ell}$ functions are defined to be real, non-singular functions.
The singularity behavior of the transition form factors $h_{j,\ell m_\ell}$ can be found by substituting Eq.~\eqref{eq:AProj.ALD} into the on-shell expansion of Eq.~\eqref{eq:H_on-shell} and equating it to its form factor decomposition
\begin{equation}
    h_{j,\ell }(s_f,Q^2) = \Mc_\ell(s_f) a_{j,\ell }(s_f,Q^2)\,,
\end{equation}
showing that the transition form factors posses a branch cut at threshold.

Given the form factor expansion of the $1+\Jc\to 2$ amplitude, we now turn our attention to the on-shell expansion of Eq.~\eqref{eq:T21_1B_sinpart}
\begin{equation}\label{eq:AProj.CC1}
    \Cc_1(P_s) = i\Hc^A_{\ell m_\ell}(P_f,P_s) i\Delta(P_s) iw^B(P_s,P_i)\,.
\end{equation}
We decompose $w$ as before, and follow a similar strategy for the off-shell amplitude $\Hc$ 
\begin{equation}\label{eq:AProj.HLDoff}
    i\Hc^A_{\ell m_\ell}(P_f,P_s)
    =  \sum_j K^A_{j,\ell m_\ell}(P_f,P_s) ih_{j,\ell }(-q_f^2,s_f,s)\,,
\end{equation}
where $s_f=P_f^2$ and $s=P^2_s$. This expansion differs from the expression in Eq.~\eqref{eq:AProj.HLD} in that $s$ is not necessarily equal to $m^2$. Again we can observe that these off-shell form factors reduce to their on-shell counterparts once the momentum $P_s$ is on-shell. This implies that the on-shell expansion of the transition form factors, for arbitrary values of $s$, take the form
\begin{equation}\label{eq:AProj.hjonexp}
        ih_j(-q_f^2,s_f,s)=ih_j(-q_f^2,s_f) + [ih_j(-q_f^2,s_f,s)]\delta\,,
\end{equation}
where the $\delta$ operator acts in the same manner as the one defined in Eq.~\eqref{eq:APRoj.ifexp}. This allows us to define the on-shell projected amplitude
\begin{equation}
    i\Hc^{A}_{{\rm on},\ell m_\ell} (P_f,P_s)
    \equiv \sum_j K^A_{j,\ell m_\ell}(P_f,P_s) ih_{j,\ell}(-q_f^2,s_f)\,,
\end{equation}
for arbitrary values of $P_s$.
We can now substitute the expansion of Eq.\eqref{eq:AProj.hjonexp} into Eq.~\eqref{eq:AProj.HLDoff} to split the singular part and the smooth functions of $s$ in Eq.\eqref{eq:AProj.CC1} 
\begin{align}
   \Cc_1(P_s)
    &= \sum_{j,l} K^A_{j,\ell m_\ell}(P_f,P_s) ih_{j,\ell}(-q_f^2) iD(P_s^{2})  K_l^B(P_f,P_s)if_l(-q_i^2) + \delta\Cc_1(P_s)\\
    &= i\Hc_{{\rm on},\ell m_\ell}^{A}(P_f,P_s) iD(P_s^{2})  iw_{\rm on}^{B}(P_s,P_i) + \delta\Cc_1(P_s)\,.
\end{align}
The kernel $\delta\Cc_1$ does not possess any singularities in the variable $P_s$ within the kinematic region of interest. This smooth kernel corresponds to $\Tb_{0|R}$ in Eq.~\eqref{eq:T21_1B_sinpart} once the dependence on the final state angular momenta has been exchanged by that of the relative momenta between the final state particles. This last equation, together with the definition of $\Hc_{\rm on}$ are the main results of this section. We complement this appendix by providing some examples for the non-trivial case of a vector current, and the implications of the constraint imposed by gauge invariance.

%%%%%%%%%%%%%%%%%%%%%%%%%%%%%%%%%%%%%%%%%%%
%  Subsection: Gauge invariance
%%%%%%%%%%%%%%%%%%%%%%%%%%%%%%%%%%%%%%%%%%%
\subsection{Examples and gauge invariance constraints}

The simplest non trivial example is that of the $1+\Jc^\mu\to 1$ electromagnetic transition between two single hadron states, both of mass $m$. The most general decomposition is given by
\begin{equation}\label{eq:AProj.1g1LD}
    \mel{p'}{\Jc^\mu(0)}{p}=(p'+p)^\mu f_1(Q^2) + (p'-p)^\mu f_2(Q^2)\,,
\end{equation}
where $\Jc^\mu(0)$ is the electromagnetic current.
Gauge invariance imposes a constraint via the relation
\begin{equation}
    (p'-p)_\mu  \mel{p'}{\Jc^\mu(0)}{p}=0\,,
\end{equation}
which implies that $f_2(Q^2)=0$. 

Whether $f_2$ is fixed to be zero for off-shell kinematics or not, depends on the chosen on-shell expansion prescription. Following the prescription described above, the on-shell projected kernel is given by Eq.~\eqref{eq:WT.wonLD}, but repeated here for clarity,
\begin{equation}\label{eq:AProj.wmuon}
    w^\mu_{\rm on}(p',p)=(p'+p)^\mu f_1(Q^2)\,,
\end{equation}
for momenta $p'$ and $p$ that need not be on the mass shell. An alternative prescription could require gauge invariance to be satisfied by the on-shell projected kernels, in which case there could be various options, one of them being
\begin{equation}\label{eq:Aproj.altwon}
    \widetilde{w}^\mu_{\rm on}(p',p)=\qty((p'+p)^\mu + \frac{p^{\prime 2}-p^2}{Q^2}(p'-p)^\mu) f_1(Q^2)\,.
\end{equation}
This prescription satisfies the requirement described in Eq.~\eqref{eq:AProj.wonreq}, and it only requires input from the physical form factor $f_1$, so in principle they could be chosen for the on-shell projection. However, it differs from our prescription because it effectively lets $f_2$ have off-shell dependence
\begin{align}
\widetilde{w}^\mu_{\rm on}(p',p)&=(p'+p)^\mu f_1(Q^2)
+(p'-p)^\mu f_2(Q^2,p^{\prime 2},p^2)\,,\\
    f_2(Q^2,p^{\prime 2},p^2)&= \frac{p^{\prime 2}-p^2}{Q^2} f_1(Q^2)\,.
\end{align}
Note that using this or any equivalent prescription would modify the function $\delta\Cc_0$ in Eq.~\eqref{eq:AProj.C0onshellexp}, nonetheless it will remain a smooth function of the variable $P_s$.

A benefit of using Eq.~\eqref{eq:AProj.wmuon} is that it provides a relationship between different pieces of the on-shell representation of the Compton-like amplitudes, see Eq.~\eqref{eq:WT.qdflg} and Eq.~\eqref{eq:app.wardrel}. This turns out to be convenient in order to explore the low-photon behavior of the amplitudes.
This allowed us, for instance, to recover the well-known Thomson amplitude in App.~\ref{sec:App.Loren}, without needing to search for all the possible tensor structures that could describe $\Tc_{11}$ and satisfy gauge invariance.
Similarly these kind of relations can prove useful when studying the finite-volume effects on matrix elements that are calculated with lattice QCD, see Ref.~\cite{Briceno:2019nns} for an example of this. 

Only for the case of gauge invariant $1+\Jc\to 2$ subprocesses we need to apply a prescription similar to Eq.~\eqref{eq:Aproj.altwon} in order to ensure that the on-shell overall amplitude $\Tc_{21}$ satisfies the Ward identity \emph{and} that the short range kernel $\Bc_{21}$ is an analytic function of the current virtuality. If $\Hc_{\rm on}$ did not satisfy the Ward identity for arbitrary kinematics, an extra term would appear in Eqs.~\eqref{eq:qfTsimp}~and~\eqref{eq:WT.qiTsimp}. However, for an arbitrary prescription of $\Hc_{\rm on}$, this extra term could be finite in the limit of vanishing current momentum $q\to0$, implying that the short range kernel $\Bc_{21}$ has a pole of the form $1/q^2$. In order to avoid this we require the prescription of $\Hc_{\rm on}$ to satisfy the Ward identity for arbitrary values of the single hadron state momentum.

There are four linearly independent tensors that can be used to decompose this amplitude. However, one of these terms picks up an extra minus sign under a parity transformation. Hence, the tensors that can be used depend on the intrinsic parity of the hadrons involved in the reaction. We separate them according to their parity behavior to get
\begin{align}
\mel{P_f,p'}{\Jc^\mu(0)}{P_i}_+&=(P_f+P_i)^\mu h_1(s_f,t,Q^2)+
     p^{\prime \mu} h_2(s_f,t,Q^2) + 
    (P_f-P_i)^\mu h_3(s_f,t,Q^2)\,, \label{eq:AProj.H+LD}\\
    \label{eq:AProj.H-LD}
\mel{P_f,p'}{\Jc^\mu(0)}{P_i}_-&=\epsilon^{\mu\nu\sigma\rho}P_{f,\nu}p'_\sigma P_{i,\rho} h_-(s_f,t,Q^2)\,,
\end{align}
where the $\pm$ subscript indicates the value of the intrinsic parity of the three hadrons respectively, e.g.\ the first amplitude is appropriate for three scalars whereas the second describes a reaction with three pseudoscalars. Here we use the Lorentz scalar $t=(p'-P_i)^2$ to characterize the behavior of the scalar energy-dependent transition form factors $h_i$.

The amplitude of Eq.~\eqref{eq:AProj.H-LD} satisfies gauge invariance without further constraints, i.e.\ we can construct the on-shell projected amplitude directly in terms of $h_-$,
\begin{equation}\label{eq:AProj.H-on}
    \Hc_{-{\rm on}}^\mu(P_f,P_i) =
    \epsilon^{\mu\nu\sigma\rho}P_{f,\nu}p'_\sigma P_{i,\rho} \,h_-(s_f,t,Q^2)\,.
\end{equation}
For the positive parity amplitude, when the external legs are on shell, the relationship
\begin{align}\label{eq:Aproj.giH+}
0=&(P_f^2 -P_i^2)  h_1(s_f,t,Q^2)  + 
\qty(P_f\cdot p'- P_i\cdot p')h_2(s_f,t,Q^2)
-Q^2h_3(s_f,t,Q^2)\\
=&(P_f^2 -m^2)  h_1(s_f,t,Q^2)  + 
\frac{s_f+t-2m^2}{2}h_2(s_f,t,Q^2)
- Q^2h_3(s_f,t,Q^2) \,,
\end{align}
must exist between the different form factors. In other words, the gauge invariant matrix element is equal to
\begin{multline}
\mel{P_f,p'}{\Jc^\mu(0)}{P_i}_+=\qty((P_f+P_i)^\mu + \frac{P_f^2-m^2}{Q^2}(P_f-P_i)^\mu ) h_1(s_f,t,Q^2)\\+
     \qty(p^{\prime \mu} + \frac{s_f+t-2m^2}{2Q^2}(P_f-P_i)^\mu ) h_2(s_f,t,Q^2)\,.
\end{multline}

In order to satisfy our prescription we will let the transition form factor $h_3$ have off-shell dependence on the incoming hadron momentum. This cannot change any physical observable because when contracted with a photon wave-function or with another tensor associated with the emission of the virtual photon, which itself satisfies the Ward identity, this factor will always vanish. Therefore the explicit form of the on-shell projected transition will be given by
\begin{multline}\label{eq:AProj.H+on}
\Hc_{+{\rm on}}^\mu(P_f,p';P_i) = \qty((P_f+P_i)^\mu + \frac{P_f^2-P_i^2}{Q^2}(P_f-P_i)^\mu)h_1(s_f,t,Q^2)\\
+\qty(\,p^{\prime \mu} + \frac{s_f+t-m^2-P_i^2}{2Q^2}(P_f-P_i)^\mu)h_2(s_f,t,Q^2)\,,
\end{multline}
for arbitrary values of the initial hadron momentum. For the remainder of this section we will assume that $P_i^2$ is in general different from $m^2$. The main results of this section are Eqs.~\eqref{eq:AProj.H-on}~and~\eqref{eq:AProj.H+on}.

To further describe this amplitude we will explicitly compute the two lowest partial waves of the final state. Each partial wave amplitude can in principle be obtained from the projection with the corresponding spherical harmonic
\begin{equation}\label{eq:Aproj.inteexpa}
    \Hc_{+{\rm on},\ell m_\ell}^\mu(P_f,P_i)  =
    \int \frac{\diff \hat{\vec{p}}^{\prime\star}}{\sqrt{4\pi}} Y^*_{\ell m_\ell}(\hat{\vec{p}}^{\prime\star}) \Hc_{+{\rm on}}^\mu(P_f,p';P_i) \,,
\end{equation}
where $\hat{{\vec{p}}}^{\prime\star}$ is the direction of the spatial part of the vector $p'$ in the final state CM frame, this frame is conventionally defined such that $P_i$ points toward the positive $z$-axis.
Because of the rotational properties of the tensors in Eq.~\eqref{eq:AProj.H+on}, and the implicit dependence on $P_i$ to define the $z$-axis, Eq.~\eqref{eq:Aproj.inteexpa} does not provide the most practical route to expand the form factors into the contributions of each partial wave.
Instead, to extract each partial wave it is more convenient to first find a Lorentz decomposition of the partial wave of interest in terms of unknown form factors $h_{j,\ell}$.
Then the orthogonality of the different helicity virtual photon wavefunctions $\epsilon(q,\lambda)$ can be exploited to generate a system of equations to solve for each of the $h_j$ form factors.

In the final state CM frame, with $P_i$ pointing towards the positive $z$-axis, the virtual photon momentum and wavefunctions are equal to
\begin{align}
q^\mu &= \frac{1}{2\sqrt{s_f}}\qty(s_f -Q^2-P_i^2,0,0,-\lambda^{1/2}(s_f,-Q^2,P_i^2))^\mu\,,\\ 
\epsilon^\mu(q,0) &= \frac{1}{2\sqrt{q^2 s_f}}\qty (\lambda^{1/2}(s_f,-Q^2,P_i^2), 0,0, -(s_f -Q^2-P_i^2))^\mu\,\,,\\ 
\epsilon^\mu(q,\pm1) &= \frac{1}{\sqrt{2}}\qty(0,\pm1,-i,0)^\mu\,,
\end{align}
where we have written every component in a Lorentz invariant fashion and used the K\"all\'en triangle function
\begin{equation}
\label{eq:KLtrian}
    \lambda(a,b,c)=a^2+b^2+c^2-2ab-2ac-2bc\,.
\end{equation}
The conservation of the azimutal component of angular momentum enforces that the transition amplitude is related to the partial wave transitions in the following manner
\begin{equation}\label{eq:AProj.pwwfproj}
     \epsilon_\mu (q,-\lambda)\Hc_{+{\rm on}}^\mu(P_f,p';P_i)
     =\sum_{\ell=\abs{\lambda}}^\infty \sqrt{4\pi} Y_{\ell \lambda}(\hat{\vec{p}}^{\prime\star})   \epsilon_\mu (q,-\lambda)\Hc_{+{\rm on},\ell \lambda}^\mu(P_f,P_i)\,,
\end{equation}
where the negative of the photon helicity corresponds to the azimutal component of angular momentum because we defined the final state CM frame with the photon traveling towards the negative $z$-axis. To study the lowest two partial waves we write down the most general Lorentz decomposition for the $S$-wave and $P$-wave amplitudes. For the $S$-wave amplitude there are two linearly independent vectors, while for the $P$-wave there are three, including the vector wavefunction describing the polarization of the final state, $\epsilon(P_f,m_\ell)$. After imposing gauge invariance one finds,
\begin{align}\label{eq:Aproj.H00LD}
    \Hc_{+{\rm on},00}^\mu(P_f,P_i)
    &= \qty( (P_f+P_i)^\mu 
    +\frac{P_f^2-P_i^2}{Q^2} (P_f-P_i)^\mu )h_{1,0}(s_f,Q^2)\,,\\
  \Hc_{+{\rm on},1 m_\ell}^\mu(P_f,P_i)
    &=   \qty( (P_f+P_i)^\mu 
    +\frac{P_f^2-P_i^2}{Q^2} (P_f-P_i)^\mu ) [\epsilon(P_f,m_\ell)^*\cdot P_i] h_{1,1}(s_f,q^2) \nn \\ \label{eq:Aproj.H1MLD}
    &\qquad \qquad + \qty(\epsilon^\mu(P_f,m_\ell)^* -\frac{[\epsilon(P_f,m_\ell)^*\cdot P_i] }{Q^2} (P_f-P_i)^\mu )h_{2,1}(s_f,q^2)\,, 
\end{align}
where the final-state vector wavefunctions in their CM frame are simply given by
\begin{equation}
\epsilon^\mu(P_f,0)=(0,0,0,1)^\mu\,,\quad \epsilon^\mu(P_f,\pm1)=(0,\mp1,-i,0)^\mu/\sqrt{2}\,.
\end{equation}

A validation of this decomposition is found by noting that the $P$-wave expansion is equivalent to what was found in \cite{Dudek:2006ej} to describe the vector-scalar electromagnetic transition, i.e.\ it can be described in terms of two independent form factors.
The behavior of higher partial waves follow a similar pattern to that of $\ell=1$. The key difference is that the vector wavefunction needs to be replaced with the one describing the angular momentum of the final state.

We now substitute Eq.~\eqref{eq:AProj.H+on} into the left hand side of Eq.~\eqref{eq:AProj.pwwfproj}, and Eqs.~\eqref{eq:Aproj.H00LD}~and~\eqref{eq:Aproj.H1MLD} into the right hand side of it.
Out of the three virtual photon helicities only two contractions are linearly independent, the $\epsilon_\mu (q,\pm\lambda)\Hc_{+{\rm on}}^\mu$ amplitudes are related by parity inversion, so that we get two equations to solve for the form factors $h_1$ and $h_2$ in terms of $h_{i,\ell}$.
Following this procedure we find for the lowest two partial-waves
\begin{align}
    h_1(s,t,Q^2) &= P_0(\cos(\theta^\star))\qty(h_{1,0}(s_f,Q^2)- \frac{\sqrt{3}}{2}\frac{\sqrt{s_f}}{\lambda^{1/2}(s_f,m^2,m^2)}h_{2,1}(s_f,Q^2))\\\nn
    &\hspace{3cm}-\sqrt{3}P_1(\cos(\theta^\star))\frac{\lambda^{1/2}(s_f,-Q^2,P_i^2)}{2\sqrt{s_f}} h_{1,1}(s_f,Q^2) + \dots\,,\\
    h_2(s,t,Q^2) &= \sqrt{3}P_1^\prime(\cos(\theta^\star))\frac{2\sqrt{s_f}}{\lambda^{1/2}(s_f,m^2,m^{2})}h_{2,1}(s_f,Q^2)+\dots\,,
\end{align}
where $P_\ell$ is the $\ell$-th Legendre polynomial and $P^\prime_\ell$ is its derivative, while $\theta^\star$ is the angle between vectors $P_i$ and $p'$ in the final state CM frame.
As expected, from the non-trivial rotational properties of the kinematic tensors chosen in the decomposition of $\Hc_{\rm on}$, the expansion of $h_1$ and $h_2$ is not given in terms of a single form factor $h_{i,\ell}$ per $\ell$ value.
In order to obtain an expansion where only one type of term contributes to $h_1$ it would be necessary to rewrite the Lorentz decomposition of $\Hc_{\rm on}$ in terms of a vector orthogonal to the $\pm 1$ photon-helicity wavefunctions, like $P_f+P_i$, and another one orthogonal to the zero helicity wave function, e.g.\
\begin{equation}
p^{\prime\mu}-\qty(1+\frac{(s_f-Q^2-P_i^2)(2t+s_f-2m^2-P_i^2+Q^2)
}{\lambda(s_f,-Q^2,P_i^2)})\frac{P_f^\mu}{2}\,.
\end{equation}
To derive this kinematic factor we have written the cosine of the angle between $P_i$ and $p'$ in terms of Lorentz scalars,
\begin{equation}
\cos\theta^\star = \frac{s_f(2t+s_f-2m^2-P_i^2+Q^2)}{\lambda^{1/2}(s_f,m^2,m^2)\lambda^{1/2}(s_f,-Q^2,P_i^2)}\,.
\end{equation}
Similarly, one would require the same orthogonality properties with the photon wavefunctions for the Lorentz decomposition of each of the partial-wave projected amplitudes $\Hc_{{\rm on},\ell m_\ell}$. This would ensure that the expansion of each of the $t$-dependent form factors $h_i$ only contains a single $h_{i,\ell}$ per $\ell$ value.

%%%%%%%%%%%%%%%%%%%%%%%%%%%%%%%%%%%%%%%%%%%
%  Section: Common Identities
%%%%%%%%%%%%%%%%%%%%%%%%%%%%%%%%%%%%%%%%%%%
\section{Common Loop Identities}
\label{sec:App.Iden}

Here we list common identities that are useful in performing the on-shell derivations of interest. The first identity relates to how the initial state interactions (ISI) of two particles manifest as a dressing over a generic smooth kernel $\Ob_{0}$ 
\begin{equation}
\label{eq:O_ISI}
    i\Ob_\text{ISI}(p',p)
   \equiv
i\Ob_{0} (p',p)
+
\xi \int\frac{\diff^{4}k}{(2\pi)^{4}}
i\Ob_{0} (p',k) 
\Delta^{(2)}(P,k)
i\Mc (k,p)\,,
\end{equation}
where the particles in the final state carry momenta $p'$ and $P-p'$, and $p$ and $P-p$ are the momenta of the initial state hadrons. 
As it was shown in \cite{Briceno:2020vgp} we can define a smooth kernel $\Ob_{j}$ with $j>0$ recursively via the loop identity
\begin{equation}\label{eq:App.loop}
\xi \int\frac{\diff^{4}k}{(2\pi)^{4}}
i\Ob_{j} (p',k)
\Delta^{(2)}(P,k) i\Kc_0 (k,p) 
= i\Ob_{j+1} (p',p) + 
\sum_{\ell, m_\ell}
i\Ob_{j,\ell m_\ell} (p)\, \rho\, i\Kc_{0,\ell}(p)\,,
\end{equation}
where the $j$-th kernel was partial-wave projected 
after intermediate particles are placed on-shell, defined via
\begin{equation}
\Ob_{j} (p',k)
=
\sum_{\ell, m_\ell}
\sqrt{4\pi}
\Ob_{j,\ell m_\ell}(p') Y^*_{\ell m_\ell}(\hat{\vec{k}}^{\star})\,,
\end{equation}
where the $k$ momentum is assumed to be on-shell.
By use of this relation we can show that the initial state interactions generate the analytic structure
\begin{align}
    i\Ob_\text{ISI}(p',p) &=
    \sum_{j=0}^\infty i\Ob_{j}(p',p)
    + \sum_{\ell, m_\ell}  \sum_{j=0}^\infty i\Ob_{j,\ell m_\ell}(p')
    \, \rho\, i\mathcal{M}_\ell(p)\\
    \label{eq:App.ISIfin}
    &=i\Ob_{\infty}(p',p)
    + \sum_{\ell, m_\ell}
    i\Ob_{\infty,\ell m_\ell}(p')
    \, \rho\, i\mathcal{M}_\ell(p)\,,
\end{align}
where we defined the smooth kernel $\Ob_{\infty}$ in the last relation.
In the case of final state interactions (FSI) the derivation follows a similar pattern
\begin{align}
    i\Ob_\text{FSI}(p',p)
   &\equiv
i\Ob_{0|n} (p',p)
+
\xi \int\frac{\diff^{4}k}{(2\pi)^{4}}
i\Mc (p',k) \Delta^{(2)} (P,k)  i\Ob_{0|n} (k,p)\\
&=i\Ob_{\infty|n}(p',p)
    + \sum_{\ell, m_\ell}i\mathcal{M}_\ell(p')
    \, \rho\,
    i\Ob_{\infty|n,\ell m_\ell}(p)\,,
    \label{eq:App.FSIfin}
\end{align}
here the label $n$ denotes that this kernel could have dressings in the initial state as well, a similar notation could be used in Eq.~\eqref{eq:App.ISIfin} to distinguish between initial and final state dressings.

The third identity that will be useful will be that of the interaction that arises from an intermediate two-particle state between two smooth kernels $\Ab_{0}$ and $\Bb_{0}$
\begin{multline}\label{eq:App.MSI}
    [i\Ab_{0}\, i\Bb_{0}]_\mathrm{MSI}(p',p)
    \equiv
\xi \int\frac{\diff^{4}k}{(2\pi)^{4}}
i\Ab_{0} (p',k) 
\Delta^{(2)}(P,k)
i\Bb_{0} (k,p)  \\
+
\xi \int\frac{\diff^{4}k}{(2\pi)^{4}}
\xi \int\frac{\diff^{4}k'}{(2\pi)^{4}}
i\Ab_{0} (p',k)
\Delta^{(2)}(P,k)
i\Mc(k,k')
\Delta^{(2)}(P,k')
i\Bb_{0} (k',p)\,.  
\end{multline}
We can use the definition of the FSI kernel to simplify this expression to
\begin{equation}
    \label{eq:App.MSI-FSI}
    [i\Ab_{0}\, i\Bb_{0}]_\mathrm{MSI}(p',p)
    \equiv
\xi \int\frac{\diff^{4}k}{(2\pi)^{4}}
i\Ab_{0} (p',k) \Delta^{(2)}(P,k) i\Bb_\text{FSI} (k,p)\,.
\end{equation}
In this case we need a generalization of the loop relation of Eq.~\eqref{eq:App.loop} for two arbitrary different kernels
\begin{equation}\label{eq:App.Ijdefin}
    \xi \int\frac{\diff^{4}k}{(2\pi)^{4}}
i\Ab_{j} (p',k) 
\Delta^{(2)}(P,k)
i\Bb_0 (k,p)
= i \Ib_{j+1}(p',p) + \sum_{\ell, m_\ell}
i\Ab_{j,\ell m_\ell} (p')
\, \rho \,
i\Bb_{0,\ell m_\ell}(p)\,,
\end{equation}
where $\Ab_{j}$ satisfies a similar recursive relation as $\Ob_{j}$ in Eq.~\eqref{eq:App.loop}, and the kernel $\Ib_j$ is smooth in the kinematic region of interest. Using identity~\eqref{eq:App.loop}, we can rewrite the second term in Eq.~\eqref{eq:App.MSI} as,
\begin{multline}
    \xi \int\frac{\diff^{4}k}{(2\pi)^{4}}
\xi \int\frac{\diff^{4}k'}{(2\pi)^{4}}
i\Ab_{0} (p',k) 
\Delta^{(2)}(P,k)
i\Mc(k,k')
\Delta^{(2)}(P,k')
i\Bb_{0} (k',p)\,
\\ =
\sum_{\ell m_\ell }i\Ab_{0,\ell m_\ell} (p')\, \rho \,
\xi \int\frac{\diff^{4}k'}{(2\pi)^{4}}
i\Mc_{\ell m_\ell}(k')
\Delta^{(2)}(P,k)
i\Bb_{0} (k',p) + 
\xi \int\frac{\diff^{4}k'}{(2\pi)^{4}}
i\Ab_{1}(p,k')
\Delta^{(2)}(P,k')
i\Bb_{0} (k',p) \\
+
\xi \int\frac{\diff^{4}k}{(2\pi)^{4}}
\xi \int\frac{\diff^{4}k'}{(2\pi)^{4}}
i\Ab_{1} (p',k) 
\Delta^{(2)}(P,k)
i\Mc(k,k')
\Delta^{(2)}(P,k')
i\Bb_{0} (k',p)\,,
\end{multline}
where we used the integral equation for $\Mc$, Eq.~\eqref{eq:M_dse} in arriving at this equality.
This identity allows us to expand Eq.~\eqref{eq:App.MSI} to get
\begin{equation}
        \label{eq:App.MSIiter}
        [i\Ab_{0}\, i\Bb_{0}]_\mathrm{MSI}(p',p)
        =
        i\Ib_1(p',p)
        +\sum_{\ell, m_\ell}
i\Ab_{0,\ell m_\ell} (p')
\, \rho \,
i\Bb_{\text{FSI},\ell m_\ell}(p) 
+ \xi \int\frac{\diff^{4}k}{(2\pi)^{4}}
i\Ab_{1} (p',k) 
\Delta^{(2)}(P,k)
i\Bb_\text{FSI} (k,p)\,.
\end{equation} 
We can iteratively apply Eqs.~\eqref{eq:App.MSI-FSI}~and~\eqref{eq:App.MSIiter} and the result of Eq.~\eqref{eq:App.FSIfin} to obtain
\begin{align}
\label{eq:App.MSIsol}
        [i\Ab_{0}\, i\Bb_{0}]_{\mathrm{MSI}}
        &=
    \sum_{j=1}^\infty
    i\Ib_j(p',p)
    +\sum_{\ell, m_\ell} i\Ab_{\infty,\ell m_\ell}(p')\,\rho\, i\Bb_{\mathrm{FSI},\ell m_\ell}(p)\,.
\end{align}

%%%%%%%%%%%%%%%%%%%%%%%%%%%%%%%%%%%%%%%%%%%
%  Section: Analytic Continuation
%%%%%%%%%%%%%%%%%%%%%%%%%%%%%%%%%%%%%%%%%%%
\section{Analytic continuation of $\Wc_{\df}$ and $\Gc$ in $s_f$}
\label{sec:App.AC}

From the form of $\Wc_{\df}$ shown in Eq.~\eqref{eq:Wdf_def} it can be seen that this amplitude will contain branch cuts in both $s_{f}$ and $s_{i}$ due to the $\Mc$ amplitudes which appear on either side and the analytic behavior of the triangle function $\Gc$. Here we show the analytic continuation of the amplitude $\Wc_{\df}$ to the second Riemann sheet of the variable $s_{f}$, while leaving variable $s_{i}$ in the physical sheet, as is required for deriving the analytic form for $t_{1\to R}$ shown in Sec.~\ref{sec:AnalyticContinuation}. This naturally also leads to a definition for the triangle function, $\Gc$, when it has also been continued to the second sheet in $s_{f}$ only. For simplicity we assume the case of a scalar current. 

We begin by finding the discontinuity of $\Wc_{\df}$ across the real $s_{f}$-axis where the discontinuity is defined to be
\begin{align}
    \label{eq:disc}
    {\rm Disc}_{x_1} f(x_1,\dots,x_n)
    =
    \lim_{\epsilon\to 0^{+}}
    \left[
    f(x_1+i\epsilon,\dots,x_n) - f(x_1-i\epsilon,\dots,x_n)
    \right]\,.
\end{align}
Applying this to the definition for $\Wc_{\df}$ given in Eq.~\eqref{eq:Wdf_def} and rearranging terms we find,
\begin{multline}
    {\rm Disc}_{s_{f}} \Wc_{\df}(s_{f},Q^{2},s_{i})
    =
    \lim_{\epsilon\to0^{+}}
    \left[
    (\Mc(s_{f,+})-\Mc(s_{f,-}))
    \Ac_{22}(s_{f,+},Q^{2},s_{i})\Mc(s_{i})
    +
    \Mc(s_{f,+})f(Q^{2})\Gc(s_{f,+},Q^{2},s_{i})\Mc(s_{i})\right.\\
    \left.-
    \Mc(s_{f,-})f(Q^{2})\Gc(s_{f,-},Q^{2},s_{i})\Mc(s_{i})
    \right]\,,
\end{multline}
where we have introduced the notation $s_{f,\pm}\equiv s_{f}\pm i\epsilon$. This may be further simplified by use of the unitarity condition for $\Mc$ which can be written as,
\begin{align}
 \Mc(s_{f,+})-\Mc(s_{f,-})
 =
 2i\rho(s_{f,+})\Mc(s_{f,-})\Mc(s_{f,+})\,,
\end{align}
as well as the Schwartz reflection principle which states,
\begin{align}
    \Mc^{*}(s_{f,+})
    =
    \Mc(s_{f,-})\,,
\end{align}
where the ``*'' superscript designates complex conjugation.
Thus, we find that the discontinuity across the $s_{f}$ axis can be written as,
\begin{align}
    {\rm Disc}_{s_{f}} \Wc_{\df}(s_{f},Q^{2},s_{i})
    =
    2i\rho(s_{f})\Mc^{*}(s_{f})\Wc_{\df}(s_{f},Q^{2},s_{i})
    +
    \Mc^{*}(s_{f})f(Q^{2}){\rm Disc}_{s_{f}}\Gc(s_{f},Q^{2},s_{i})\Mc(s_{i})\,.
\end{align}

Analytically continuing through the $s_{f}$ branch cut gives the relations,
\begin{align}
    \Mc(s_{f,-}) &= \Mc^{\rm II}(s_{f,+})\,,\\
    \Wc_{\df}(s_{f,-},Q^{2},s_{i}) &= \Wc_{\df}^{\rm II,I}(s_{f,+},Q^{2},s_{i})\,,
\end{align}
where the superscripts on the amplitudes on the right hand side tell us the variable $s_{f}$ is now on the second sheet. Using these we may re-write the discontinuity relation as,
\begin{multline}
    \Wc_{\df}(s_{f,+},Q^{2},s_{i})-\Wc_{\df}^{\rm II,I}(s_{f,+},Q^{2},s_{i})
    =
    2i\rho(s_{f,+})\Mc^{\rm II}(s_{f,+})\Wc_{\df}(s_{f,+},Q^{2},s_{i})\\
    +
    \Mc^{\rm II}(s_{f,+})f(Q^{2}){\rm Disc}_{s_{f}}\Gc(s_{f},Q^{2},s_{i})\Mc(s_{i})\,,
\end{multline}
where the limit as $\epsilon\to0^{+}$ has been left implicit.
Solving this for $\Wc^{\rm II,I}_{\df}$ we find,
\begin{align}
    \Wc^{\rm II,I}_{\df}(s_{f},Q^{2},s_{i})
    &=
    (1-2i\rho(s_{f})\Mc^{\rm II}(s_{f}))\Wc_{\df}(s_{f},Q^{2},s_{i})
    -
    \Mc^{\rm II}(s_{f})f(Q^{2}){\rm Disc}_{s_{f}}\Gc(s_{f},Q^{2},s_{i})\Mc(s_{i})\nn\\
    &=
    \Mc^{\rm II}(s_{f})
    (\Ac_{22}(s_{f},Q^{2},s_{i}) + f(Q^{2})\Gc^{\rm II,I}(s_{f},Q^{2},s_{i}))
    \Mc(s_{i})\,,
\end{align}
where to go from the first to second line we used the following relation between the first and second sheet of the scattering amplitude,
\begin{align}
    \Mc^{\rm II}(s_{f})
    =
    \frac{1}{1+2i\rho(s_{f})\Mc(s_{f})}\Mc(s_{f})\,,
\end{align}
and $\Gc^{\rm II,I}$ is defined to be,
\begin{align}
    \Gc^{\rm II,I}(s_{f},Q^{2},s_{i})
    =
    \Gc(s_{f},Q^{2},s_{i})
    -
    {\rm Disc}_{s_{f}}\Gc(s_{f},Q^{2},s_{i})\,.
\end{align}

Finally, we wish to find an explicit form for ${\rm Disc}_{s_{f}}\Gc$. For the case of $S$-wave in both initial and final partial waves with a scalar current, this can be done using the form for the singularity structure of $\Gc$ given in Eq.~(A44) of Ref.~\cite{Briceno:2020vgp} which we show here for convenience,
\begin{align}
    \label{eq:Sing_G}
    {\rm Sing}\ \Gc(s_{f},Q^{2},s_{i})
    =
    \frac{i}{16\pi\lambda^{1/2}(s_{f},-Q^{2},s_{i})}
    \left[
    \log\left(
    \frac{\rho(s_{f})+b_{f}}
    {\rho(s_{f})-b_{f}}
    \right)
    +
    \log\left(
    \frac{\rho(s_{i})+b_{i}}
    {\rho(s_{i})-b_{i}}
    \right)
    \right]\,,
\end{align}
where 
\begin{align}
    b_{f}
    &=
    \frac{Q^{2}+s_{f}-s_{i}+2(m_{1}^{2}-m_{2}^{2})(1-(Q^{2}+s_{f}+s_{i})/(2s_{f}))}
    {16\pi\xi^{-1}\lambda^{1/2}(s_{f},-Q^{2},s_{i})}\,,
\end{align}
where the particle with mass $m_{1}$ couples to the current while the particle with mass $m_{2}$ acts as a spectator and $\lambda$ is defined in Eq.~\eqref{eq:KLtrian}. There is a similar definition for $b_{i}$ but with each $f$ subscript replaced with $i$ and vice-versa.
The triangle function $\Gc$ is independent of whether or not the particles in the loop are identical, so that the function $b_{f/i}$ necessitates a factor of $\xi$ to compensate for the one appearing in $\rho$.
Here we are exploiting that Eq.~\eqref{eq:Sing_G} contains all of the non-analytic structure of $\Gc$, thus to calculate ${\rm Disc}_{s_{f}}\Gc$ we can apply Eq.~\eqref{eq:disc} for the variable $s_{f}$ to Eq.~\eqref{eq:Sing_G}. 

The key to calculating this quantity is to realize that as we take the limit we need not worry about possible branch cuts that may arise from $\lambda^{1/2}(s_{f},-Q^{2},s_{i})$. This can be seen by expanding the logarithms of ${\rm Sing}\ \Gc$ to show that it only depends on odd powers of $\lambda^{1/2}(s_{f},-Q^{2},s_{i})$. Thus when combined with the factor out front we only get integer powers of $\lambda(s_{f},-Q^{2},s_{i})$. Therefore, we only need to consider the behavior of the phase space factors. A consequence of choosing the branch cut in $s_{f}$ to lie along the positive real axis is that on either side of the cut $\rho(s_{f,+})=-\rho(s_{f,-})$ while $\rho(s_{i})$ has no dependence on $s_{f}$ and therefore will not pick up the same change in sign. Thus when we take the limit of these functions as they appear in Eq.~\ref{eq:Sing_G} we find,

\begin{align}
    \lim_{\epsilon\to0}\frac{1}{\lambda^{1/2}(s_{f,\pm},-Q^{2},s_{i})}
\log\left(
    \frac{\rho(s_{i})+b_{i}}{\rho(s_{i})-b_{i}}
    \right)
    &=
    \frac{1}{\lambda^{1/2}(s_{f},-Q^{2},s_{i})}
\log\left(
    \frac{\rho(s_{i})+b_{i}}{\rho(s_{i})-b_{i}}
    \right),\\
    \lim_{\epsilon\to0}\frac{1}{\lambda^{1/2}(s_{f,\pm},-Q^{2},s_{i})}
\log\left(
    \frac{\rho(s_{f,\pm})+b_{f}}{\rho(s_{f,\pm})-b_{f}}
    \right)
    &=\pm\frac{1}{\lambda^{1/2}(s_{f},-Q^{2},s_{i})}
\log\left(
    \frac{\rho(s_{f})+b_{f}}{\rho(s_{f})-b_{f}}
    \right)\,.
\end{align}

With this in mind we can then apply Eq.~\eqref{eq:disc} to Eq.~\eqref{eq:Sing_G} to calculate ${\rm Disc}_{s_{f}}\Gc$ to be
\begin{align}
    {\rm Disc}_{s_{f}}\Gc(s_{f},Q^{2},s_{i})
    &=
    \frac{i}{8\pi\lambda^{1/2}(s_{f},-Q^{2},s_{i})}
    \log\left(
    \frac{\rho(s_{f})+b_{f}}{\rho(s_{f})-b_{f}}
    \right)\,.
\end{align}

\bibliographystyle{apsrev} %%% physical review
\bibliography{bibi} %%% ref.bib file
\end{document}